\documentclass{article}
\usepackage{arxiv}
\usepackage{graphicx,amsmath,amsfonts,amscd,amssymb,pstricks}
\usepackage{color}% Include figure files
\usepackage{dcolumn}% Align table columns on decimal point
\usepackage{bm}% bold math
\usepackage{longtable}
\usepackage{mathrsfs}
\usepackage{textcomp}
\usepackage{esvect}
\usepackage{float}
\usepackage{mathtools,amsfonts}
\usepackage[USenglish,british]{babel}
\usepackage{environ}
\usepackage{xifthen}
\usepackage{xargs}		% programming: better newcommand
\usepackage{hyperref}
\usepackage{physics,mathtools}
\usepackage[separate-uncertainty = true]{siunitx}
\usepackage{multirow}
\usepackage{comment}
\usepackage{enumerate}
\usepackage{appendix}
\usepackage{tikz}	
\usetikzlibrary{arrows, calc, matrix, patterns, decorations.markings, shapes, decorations.pathmorphing, shadows.blur}
\usepackage[utf8]{inputenc}
\usepackage{enumitem}

\newcommand{\av}[1]{\left<{#1}\right>}
\newcommand{\ie}{\textit{i.e.}\ }

\newcommand{\eg}{\textit{e.g.}\ }

\newcommand{\wrt}{\textit{w.r.t.}\ }

\newcommand{\rhs}{r.h.s.\ }

\newcommand{\ham}{\mathcal{H}}

\newcommand{\eps}{\varepsilon}

\newcommand{\Ji}[1]{J_{#1,\text{i}}}
\newcommand{\Jf}[1]{J_{#1,\text{f}}}
\newcommand{\phii}[1]{\phi_{#1,\text{i}}}

\title{Manipulation of transverse emittances in circular accelerators by crossing non-linear 2D resonances}
\author{A.~Bazzani\\
Physics and Astronomy Department, Bologna University and INFN Bologna\\
\And
F.~Capoani\\
Physics and Astronomy Department, Bologna University and INFN Bologna\\
Beams Department, CERN, Esplanade des Particules 1, 1211 Meyrin, Switzerland\\
\And 
M. Giovannozzi\thanks{Corresponding author: massimo.giovannozzi@cern.ch}\\
Beams Department, CERN, Esplanade des Particules 1, 1211 Meyrin, Switzerland}

\begin{document}
\maketitle

\begin{abstract}%
Controlling non-linear effects in the transverse dynamics of charged particle beams in circular accelerators opens new possibilities for controlling some of the beam properties. Beam splitting by crossing a stable 1D non-linear resonance is part of the routine operation of the CERN Proton Synchrotron. The beam undergoes trapping and transport inside stable islands created in the horizontal plane to allow multi-turn extraction towards the Super Proton Synchrotron, where the beam is used for fixed-target experiments. This process acts only on the horizontal beam emittance, inducing a reduction of its initial value. In this paper, we present a generalisation of this approach, in which both transverse planes are affected by the proposed technique. We will discuss in detail how to manipulate the transverse emittances by means of a controlled crossing of a 2D non-linear resonance. The novel technique will be presented by discussing the theoretical analysis of a Hamiltonian model, as well as simulating the performance of the proposed manipulation using a more realistic non-linear symplectic map.  
\end{abstract}
%
%\PACS{05.45.-a, 29.20.-c, 29.27.Ac, 29.27.Bd, 45.20.Jj}
%\PACS{
%      {PACS-key}{discribing text of that key}   \and
%      {PACS-key}{discribing text of that key}
%     } % end of PACS codes
%} %end of abstract
%
%\maketitle
%
\section{Introduction}
\label{sec:intro}

In circular particle accelerators, the beam dynamics is governed by a number of invariants. This is the case of transverse emittances, which are preserved under the condition that radiation effects can be neglected, which is often the case for hadron accelerators at intermediate beam energy, and time-dependent effects are absent. The very existence of these invariants is related to the existence of integrals of motion of the Hamiltonian character of the beam dynamics. 

This picture, however, has to be completely reviewed whenever non-linear resonances and time-dependent effects are introduced in the system. In this case, the invariance of the transverse emittances is broken, which implies a potential harmful impact on the accelerator performance due to the diffusion of orbits in phase space~\cite{Bazzani9948,Kandrup1999DiffusionAS,NEISHTADT2006158,Neishtadt_2019}.

Perturbative theory~\cite{Nekhoroshev:1971aa} has shown that for quasi-integrable systems one can define quasi-invariant quantities whose value changes by a small amount even after extremely long times. In the presence of non-linear effects, non-linear invariants can be found, which represent the generalisation of the Courant-Snyder invariant~\cite{Courant:593259} for the case of linear dynamics. In this respect, the theory of Normal Forms (see, \eg \cite{Bazzani:262179} for an overview) provides exactly the tools to evaluate, among other observables, these invariants for a non-linear symplectic map near an elliptic fixed point.

Indeed, the lack of invariance for the transverse emittances implies a degraded accelerator performance in terms, \eg of emittance growth and particle loss. There are, however, also some advantages that we would like to highlight and explore. One could devise appropriate beam manipulations in which the transverse emittances are acted upon in a controlled way. 

In recent years, the proposal of the so-called Multi-Turn Extraction (MTE) at the CERN Proton Synchrotron (PS)~\cite{PhysRevLett.88.104801,PhysRevSTAB.7.024001,Giovannozzi:987493,PhysRevSTAB.12.014001} and its successful operational implementation~\cite{Borburgh:2137954,PhysRevAccelBeams.20.014001,PhysRevAccelBeams.20.061001,PhysRevAccelBeams.22.104002} opened the way to a novel attitude towards applications of non-linear beam dynamics. The principle of MTE relies on the adiabatic crossing of a non-linear resonance in the horizontal plane, with the goal to trap particles inside the stable islands of the phase space and then transport them towards high amplitudes prior to extraction without any beam loss. This manipulation generates a beam structure that extends beyond the length of the PS circumference, which is an essential feature of MTE. However, the beam splitting, achieved when particles are trapped inside the stable islands, induces a reduction of the horizontal emittance only, \ie the emittance of each of the five generated beamlets is smaller than that of the initial single-Gaussian beam, which is related to the 1D character of the resonance used.

We would like to extend this approach by considering the adiabatic crossing of a non-linear coupled resonance, which should provide a control of the emittances in the horizontal and vertical planes. It is worth mentioning that an intermediate step in this direction is performed by crossing the coupling resonance in the presence of linear coupling. In this case, it is well known~\cite{Metral:529690,PhysRevSTAB.10.064003,PhysRevAccelBeams.23.044003} that the transverse emittances can be exchanged between the two planes. Furthermore, a recent paper showed how the process of resonance crossing in the presence of linear coupling can be best understood by means of the Hamiltonian theory~\cite{PhysRevAccelBeams.24.094002}.

Crossing a non-linear 2D resonance provides extended capabilities in terms of control and manipulation of transverse emittances and could be pursued both from the theoretical and experimental point of view. The source of inspiration is the analysis of the impact of crossing the Walkinshaw, \ie $\omega_x-2\omega_y=0$, resonance~\cite{PhysRevLett.110.094801,chao2015emittance}, and for the $2\omega_x-\omega_y$ resonance~\cite{kallestrup:ipac2021-mopab019}. However, the focus of the analysis performed in Ref.~\cite{PhysRevLett.110.094801} was to estimate the undesired emittance growth due to the resonance crossing, whereas our aim is to intentionally manipulate the transverse emittances for specific applications.

In this paper, we propose a general approach to emittance sharing based on adiabatic invariance and separatrix-crossing theory. If the linear frequencies are slowly modulated so to cross selected resonances, the area of the phase space enclosed by the separatrix changes and particles can jump between different phase-space regions, which results in a change of their adiabatic invariant. The statistical analyses show that, starting from a Gaussian distribution of initial conditions with emittances $\eps_{x,\text{i}}$, $\eps_{y,\text{i}}$, at the end of the resonance crossing process the emittances are given by $\eps_{x,\text{f}}\propto\eps_{y,\text{i}}$ and $\eps_{y,\text{f}}\propto\eps_{x,\text{i}}$, with factors depending on the order of the crossed resonance.

The approach presented in this paper has been assessed by means of a detailed analysis of the phase-space topology of simple Hamiltonian systems, which is the basis for applying the adiabatic theory. The technique is then probed using more realistic map models. Extensive numerical simulations have been performed to determine the dependence of the results on the various system parameters.

The structure of the paper is the following: in Section~\ref{sec:theory} the models are introduced, and the phase-space topology of the Hamiltonians described is studied in detail, including considerations on some specific low-order resonances, which are useful for applications. The results of numerical simulations are presented and discussed in Section~\ref{sec:num_res}, while some conclusions are drawn in Section~\ref{sec:conc}. Moreover, a discussion on which type of magnet excites a given resonance is given in Appendix~\ref{app:magnets}, while a short digression on the motion in the resonant condition is reported in Appendix~\ref{app:rescond}.

\section{Theoretical framework} \label{sec:theory}

\subsection{General considerations}

The starting point is the choice of the model used for our analyses, which is the H\'enon-like~\cite{henon} 4D symplectic map that describes the transverse betatron motion in a FODO cell with non-linearities~\cite{Bazzani:262179}. Such a map, written in Courant-Snyder normalised co-ordinates, is composed of rotations of frequencies $\omega_x$ and $\omega_y$ and a $2(r+1)$-polar kick, \ie
\begin{equation}
\renewcommand\arraystretch{1.8}
\begin{pmatrix}
x'\\p'_x\\ y' \\p'_y
\end{pmatrix} = R(\omega_x,\omega_y) \begin{pmatrix} x &\\ p_x &+ \sqrt{\beta_x} \Re \qty[\left ( \dfrac{k_r + i j_r}{r!} \right ) \left (\sqrt{\beta_x}\, x+i\sqrt{\beta_y}\,y \right )^r] \\ y &\\ p_y &- \sqrt{\beta_y} \Im \qty[\left (\dfrac{k_r + i j_r}{r!} \right ) \left (\sqrt{\beta_x}\,x+i\sqrt{\beta_y}\,y \right )^r] \end{pmatrix}\, ,
\label{eq:henon}
\end{equation}
where $R(\omega)$ is a 2D rotation matrix and $R(\omega_x,\omega_y)=\mathrm{diag}\qty(R(\omega_x),R(\omega_y))$, while $k_r$ and $j_r$ are the normal and the skew strength of the $2(r+1)$-polar magnet, respectively. They are obtained by considering the following expression for the transverse magnetic field
\begin{equation}
 B_y +i B_x = B\rho \sum_{r=1}^{M} \left ( k_r+ij_r\right ) \frac{\left ( x+i y\right )^r}{r!}  \, ,
\end{equation}
where $B\rho$ is the beam magnetic rigidity.

In certain situations, it is interesting to introduce an explicit amplitude-detuning effect in the map of Eq.~\eqref{eq:henon} that models the case where a magnetic multipole excites the resonance, whereas the effect of other magnetic elements, not modelled as kicks in the map, is to generate an amplitude-dependent detuning. In this case, the rotation matrix in Eq.~\eqref{eq:henon} is replaced by a rotation matrix $R(\omega_x+\alpha_{xx}J_x + \alpha_{xy}J_y, \omega_y+\alpha_{xy}J_x + \alpha_{yy}J_y)$, where the linear actions $J_x=(x^2+p_x^2)/2$, and $J_y=(y^2+p_y^2)/2$ have been used, which defines an amplitude-dependent 4D rotation.

We say that $\omega_x, \omega_y$ satisfy a $(m,\, n)$ difference resonance condition if the following holds
\begin{equation}
    m\omega_x - n\omega_y = 2\pi k \qquad m, n \in \mathbb{N}, k \in \mathbb{Z} \, , 
    \label{eq:rescond}
\end{equation}
and the resonance order is given by $m+n$. 

Normal Form theory applied to the map of Eq.~\eqref{eq:henon} close to a $(m,\, n)$ resonance condition allows a resonant Normal Form to be built, from which a quasi-resonant interpolating Hamiltonian can be derived~\cite{Bazzani:262179}. The analysis focuses on the resonances of orders $3$ and $4$ that, at leading order in the actions, are possible to excite using common magnetic elements (the details about which magnet type can excite a given resonance are given in Appendix~\ref{app:magnets}) according to the following scheme
\begin{itemize}
    \item $(1,2)$ resonance: normal sextupole ($j_2=0$);
    \item $(2,1)$ resonance: skew sextupole ($k_2=0$);
    \item $(3,1)$ resonance: skew octupole ($k_3=0$);
    \item $(1,3)$ resonance: skew octupole ($k_3=0$).
\end{itemize}

We remark that the Normal Form approach provides the resonant terms due to a given non-linearity as perturbations in the actions $J_x$ and $J_y$, instead of using the resonance strength as the perturbation parameter.

We also remark that the correspondence between magnet type and resonance is valid for the case of a single kick, \ie for a map of the form of Eq.~\eqref{eq:henon}. In the case of a system with two non-linear kicks, the fourth-order resonances can also be excited by using a combination of normal and skew sextupoles.

\subsection{Phase-space topology of the Hamiltonian model}

The Normal Form Hamiltonian in the resonant case, written in action-angle variables reads
\begin{equation}
    \ham(\phi_x,\,J_x,\,\phi_y,\,J_y) = \omega_x J_x + \omega_y J_y + \alpha_{xx}J_x^2 + 2\alpha_{xy} J_x J_y + \alpha_{yy} J_y ^2 + G J_x^{m/2} J_y^{n/2}\cos(m\phi_x-n\phi_y) \, ,
    \label{eq:initial_ham}
\end{equation}
where the amplitude-detuning parameters $\alpha_{xx}$, $\alpha_{xy}$, $\alpha_{yy}$ have been introduced and the quasi-resonance condition is given by $m\,\omega_x - n\,\omega_y\approx 0$. The resonance-strength parameter $G$ is directly proportional to the magnet strength $k_r$ or $j_r$, as one can verify by computing the resonant Normal Form Hamiltonian for map~\eqref{eq:henon} using, \eg software presented in Ref.~\cite{Bazzani:1995vj}.

The canonical transformation (see~\cite[p.~410]{Arnold:937549}) 
\begin{equation}
\begin{aligned}
J_x &= mJ_1\,,            & \phi_1 &= m \phi_x -n \phi_y\,, \\ 
J_y &= J_2-nJ_1\,,         & \phi_2 &= \phi_y\,, \
\end{aligned}
\end{equation}
introduces the fast and slow phases and casts the Hamiltonian into the form
\begin{equation}
%\begin{split}
    \ham(\phi_1,J_1) =\delta J_1 +\alpha_{12} J_1 J_2 +\alpha_{11}J_1^2 +G(mJ_1)^{\frac{m}{2}}(J_2-nJ_1)^{\frac{n}{2}}\cos \phi_1 + \Big[\omega_yJ_2+ \alpha_{22}J_2^2\Big]\,,
%\end{split}
\label{eq:ham_mn_J1J2}
\end{equation}
where $\delta=m\,\omega_x-n\,\omega_y$ is the resonance-distance parameter, and the new constants $\alpha_{11}$, $\alpha_{12}$, and $\alpha_{22}$ are functions of $\alpha_{xx}$, $\alpha_{xy}$ and $\alpha_{yy}$ according to
\begin{equation}
    \begin{split}
        \alpha_{11} &= m^2\alpha_{xx} - 2m\, n\, \alpha_{xy}+n^2\alpha_{yy}\, ,\\
        \alpha_{12} &= 2(m\, \alpha_{xy}-n\, \alpha_{yy})\, ,\\
        \alpha_{22} &= \alpha_{yy}\, .
    \end{split}
\end{equation}

We remark that the term in square brackets of Eq.~\eqref{eq:ham_mn_J1J2} can be discarded as it is a function of $J_2$ only, which is an integral of motion since $\pdv*{\ham}{\phi_2}=0$. Hence, it represents a constant additive term of the Hamiltonian. Furthermore, we remark that the term $\alpha_{12}$ induces a shift in the location of the resonance crossing, which occurs for $\delta + \alpha_{12} J_2=0$, thus making the resonance-crossing process dependent on the value of $J_2$ ( a time-independent quantity). We remark also that the condition $J_y>0$ constrains the motion within the circle $J_1 < J_2/n$, which we call the \textit{allowed circle}. We remark that the existence of the allowed circle is a consequence of having chosen a difference resonance, \ie with the minus sign in Eq.~\eqref{eq:rescond}. Sum resonances do not fulfil this property that is essential for emittance exchange.

To study the phase-space structure, it is convenient to express Eq.~\eqref{eq:ham_mn_J1J2} using the rescaled variable $\tilde{J}_1 =J_1/J_2$, that gives the Hamiltonian
\begin{equation}
    \tilde{\ham}(\phi_1,\tilde{J}_1) =\frac{\delta}{G J_2^{\frac{m+n-2}{2}}} \tilde{J}_1 +\frac{\alpha_{12}}{G J_2^{\frac{m+n-4}{2}}} \tilde{J}_1 +\frac{\alpha_{11}}{G J_2^{\frac{m+n-6}{2}}} \tilde{J}_1^2 +(m\tilde{J}_1)^{\frac{m}{2}}(1-n\tilde{J}_1)^{\frac{n}{2}}\cos \phi_1\,.
\label{eq:ham_mn_rescaled}
\end{equation}
It appears that the resonance-crossing process is actually governed by the parameter
\begin{equation}
\eta = \frac{\delta}{G J_2^{\frac{m+n-2}{2}}} \, .
\label{eq:eta}
\end{equation}
Therefore, there is an interplay between the distance from resonance, $\delta$, the multipole strength, proportional to $G$, and the invariant action $J_2$. We also remark that the coefficients $\alpha_{12}, \alpha_{11}$ are rescaled by the quantity $1/(G J_2^{\frac{m+n-6}{2}})$.

The equations of motion for the Hamiltonian of Eq.~\eqref{eq:ham_mn_J1J2} are
\begin{equation}
\begin{split}
\dot{\phi}_1 &= \phantom{-}\pdv{\ham}{J_1}= 
\delta+2\alpha_{11}J_1+\alpha_{12} J_2 +\frac{m}{2}G(mJ_1)^{\frac{m}{2}-1} (J_2-nJ_1)^{\frac{n}{2} -1} \Big[ mJ_2 -n (m+n)J_1 \Big]\cos \phi_1\,,\\
\dot{J}_1 &= -\frac{\partial \ham}{\partial \phi_1}=G \, (mJ_1)^{\frac{m}{2}}(J_2-nJ_1)^{\frac{n}{2}}\sin \phi_1\,,
\end{split}
\label{eq:motion}
\end{equation}
and the phase-space topology that is originated by them depends both on $m$ and $n$, although some features do not. 

The knowledge about the existence of the fixed points of Eq.~\eqref{eq:motion} and their stability is essential for understanding the phase-space topology. The solutions of the equation $\pdv*{\ham}{\phi_1}=0$ that satisfy the condition $J_2-nJ_1=0$, are particularly relevant for our study, since they lie on the border of the allowed circle, and for this reason, these solutions have to be unstable fixed points and are computed by solving
%When unstable fixed points are present on the border of the allowed circle, the separatrix connecting them is the so-called \textit{coupling arc}. The equation $\pdv*{\ham}{J_1}=0$ gives
%
\begin{equation}
\cos\phi_1=\frac{2(\delta+2\alpha_{11} J_1+\alpha_{12} J_2)}{Gm^{m/2}\left[n^2J_1-m(J_2-nJ_1)\right]J_1^{\frac{m}{2}-1}(J_2-nJ_1)^{\frac{n}{2}-1}}\, .
\label{eq:ufp}
\end{equation}

When imposing the condition $J_1-nJ_2=0$, the \rhs of  Eq.~\eqref{eq:ufp} is not singular only if $n=1$ or $n=2$ (the exactly resonant case will be discussed later). 

The separatrix that passes through the unstable fixed points on the border of the allowed circle is called \textit{coupling arc} (as in Ref.~\cite{PhysRevLett.110.094801}), and is found by solving the equation
\begin{equation}
    \ham(\phi_1, J_1) = \delta \frac{J_2}{n} + \qty(\frac{\alpha_{11}}{n} + \frac{\alpha_{12}}{n})J_2^2\,,
\end{equation}
which can be rewritten as
\begin{equation}
    n\delta + \alpha_{11}n^2(J_2+nJ_1) + n\alpha_{12}J_2= G m^{m/2} J_1^{m/2}(J_2-nJ_1)^{\frac{n}{2}-1}\cos\phi_1\,.
    \label{eq:couplarc}
\end{equation}

For $n=1$, the term $(J_2-nJ_1)^{1/2}$ appears in the numerator of Eq.~\eqref{eq:ufp} with a positive power, and when $J_1=J_2$, \ie on the allowed circle, $\cos\phi_1=0$, so $\phi_1=\pm\pi/2$. With no amplitude detuning, the equation of the coupling arc reads
\begin{equation}
\delta(J_2-J_1)^{1/2} = Gm^{m/2} J_1^{m/2} \cos\phi_1\, ,
\end{equation}
and the existence of solutions requires $\delta \cos \phi_1 > 0$. If $\delta>0$ the coupling arc lies in the right hemicircle, while for $\delta<0$ it lies in the left one. Furthermore, for large values of $|\delta|$ the coupling arc is very close to the allowed circle, as it can be seen from the equation in the limit $|\delta| \to +\infty$.

For $n=2$, the term $(J_2-2J_1)$ disappears from the denominator of Eq.~\eqref{eq:ufp}, and the coupling-arc intersections are found for 
\begin{equation} 
\cos\phi_1 = 2^{\frac{m}{2}-1}\frac{\delta + (\alpha_{11}+\alpha_{12})J_2}{Gm^{m/2}J_2^{m/2}} \, , 
\end{equation}
which exist as long as $|\cos\phi_1|\le 1$, and they do not depend on $J_1$. In this case, in the absence of amplitude detuning, we obtain a simple expression for the coupling arc
\begin{equation} 
J_1 = \qty(\frac{\delta}{2 Gm^{m/2}\cos\phi_1})^{2/m}\, .
\label{eq:couplarc_m2}
\end{equation}

Once the Hamiltonian of Eq.~\eqref{eq:ham_mn_J1J2} is recast in Cartesian coordinates $(X=\sqrt{2J_1}\cos\phi_1,\,Y=\sqrt{2J_1}\sin\phi_1)$, one can observe that the other fixed points, which could be associated to the presence of other separatrices, can be found only on the $X$ axis due to symmetry reasons. 

First of all, we remark that  the origin $(X=0,\,Y=0)$ is a fixed point 
%
%\begin{equation} 
%\dot Y = -\frac{Gm^{m/2}}{\sqrt{2}}\qty(\frac{X^2+Y^2}{2})^{m/2 %-1/2} \, ,
%\end{equation}
%
only if $m > 1$. In this case, we can study the isoenergetic surface of the origin from the equation $\ham(\phi_1,J_1)=0$, \ie
\begin{equation} 
J_1(\delta + \alpha_{11}J_1 + \alpha_{12}J_2 + Gm^{m/2}J_1^{m/2 - 1}(J_2-nJ_1)^{n/2}\cos\phi_1) = 0 \, ,
\end{equation}
which is solved for $J_1=0$ or for 
\begin{equation}
\delta + \alpha_{12}J_2 = - \alpha_{11}J_1 - Gm^{m/2}J_1^{m/2 - 1}(J_2-nJ_1)^{n/2}\cos\phi_1 \, .
\end{equation}
%
%
%Unless $m=2$, this second surface does not pass through the origin (except in the exact resonant condition, when $\delta+\alpha_{12}J_2=0$): it therefore does not describe an allowed trajectory of a particle which crosses the origin, and the unique possibility for a point in $J_1=0$ is to lie there indefinitely.
%

For $m=2$, we can solve analytically the case without amplitude-detuning terms, as the equation becomes
\begin{equation}
\delta = - 2G(J_2-nJ_1)^{n/2}\cos\phi_1\,.
\end{equation}
A solution $J_1(\phi_1)$ that passes through the origin when $\acos(\delta/(2GJ_2^{n/2}))$ exists, \ie for $|\delta|\le 2GJ_2^{n/2}$. The solution lies in the positive-$X$ domain if $\delta<0$, and in the negative one if $\delta>0$.
For $m>2$, the origin is a genuine fixed point and the Hamiltonian can be linearised around the origin using the coordinates $X$, $Y$.
%
%\begin{equation} \pdv[2]{\ham}{X}\eval_{\substack{X=0\\Y=0}} %=\pdv[2]{\ham}{Y}\eval_{\substack{X=0\\Y=0}} =\delta\,, \qquad %\text{and}\ \qquad \pdv{\ham}{X}{Y}\eval_{\substack{X=0\\Y=0}}=0 %\qquad \text{for } m>2 \, .
%\end{equation}
%
One obtains a simple rotator Hamiltonian, \ie $\ham_\text{lin} = \delta(X^2+Y^2)/2$, which shows that the origin is an elliptic fixed point.

Finally, additional fixed points might exist on the axis $Y=0$, and they should be solutions of $\pdv*{\ham}{X}=0$, having set $Y=0$. The equation reads 
\begin{equation}
    \delta + \alpha_{11}X^2 + \alpha_{12}J_2 + \frac{G}{2}\qty(\frac{m}{2})^\frac{m}{2} \qty(J_2-\frac{n}{2}X^2)^{\frac{n}{2}-1}\,X^{m-2}\qty[2mJ_2-2-n(m+n)X^2]=0\,.
\label{eq:fixpts}
\end{equation}
The number of real solutions of Eq.~\eqref{eq:fixpts} that lie inside the allowed circle depends on the degree of the resulting polynomial in $X$, which is determined by the order of the resonance condition. Therefore, the topology of the phase space of higher-order resonances can be very complicated, and its detail is a crucial element for the feasibility of emittance sharing. A specialised discussion on fixed points on the $Y=0$ axis is carried out for each resonance taken into consideration in our study in Section~\ref{sec:resonances}.

We remark that when $\delta+\alpha_{12}J_2=0$, \ie the resonance condition is met, and $\alpha_{11}=0$, nontrivial solutions of Eq.~\eqref{eq:fixpts} are given by
\begin{equation}
    2mJ_2 -n(m+n)X^2 = 0\,, \qquad \text{or} \qquad  X = \pm \sqrt\frac{2mJ_2}{n(m+n)}\,.
\end{equation}
The two symmetrical solutions are both stable fixed points. For the origin, the previous discussion holds, having set $\delta=0$. Moreover, the coupling arc equation at resonance becomes $\cos\phi_1 =0$, and the coupling arc is reduced to the diameter of the allowed circle passing through $\phi_1 = \pm\pi/2$, for any value of $m$ and $n$. Separatrices that are not coupling arcs approximate the behaviour of a coupling arc close to the resonance (see, \eg the top-right phase-space portrait of Fig.~\ref{fig:phsp13}).

In general, at resonance, the allowed circle is symmetrically divided in two regions. Hence, whatever the resonance is crossed, if $\alpha_{11}=0$ there is always a neighbourhood of the resonant condition $\delta+\alpha_{12}J_2=0$ where the phase space is divided into two regions. This is the ideal condition to perform emittance sharing, as it will be shown in Section~\ref{sec:sharingproc}.

In the following, we analyse some resonances that can be excited using magnetic elements commonly installed in particle accelerators.

\subsection{Motion close to low-order resonances} \label{sec:resonances}
We now compute the most important features of the phase space of the resonant Normal Form Hamiltonian for low-order resonances excited by sextupole or octupole magnets. The theory of emittance sharing relies on separatrix crossing, therefore we need to know which fixed points exist in the phase space, their stability, and where separatrices exist. In general, we will search for unstable fixed points on the allowed circle, which give rise to a coupling arc, for stable fixed points on $\phi_1=0$ or $\phi_1=\pi$, and for possible extra separatrices.

\subsubsection{Resonance \texorpdfstring{$(1,2)$}{(1,2)}}

\begin{figure}
    \centering
    \includegraphics[width=.7\textwidth]{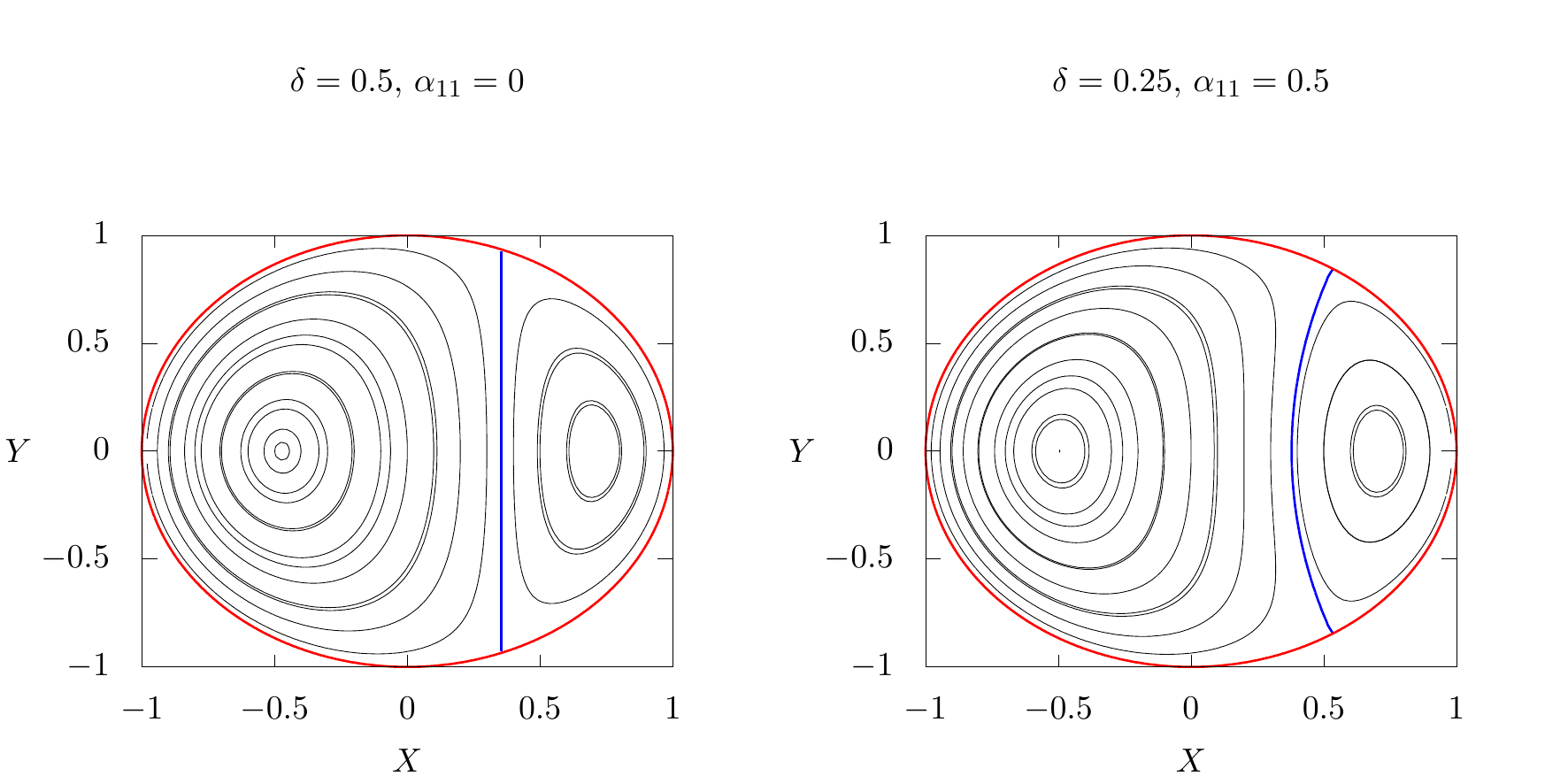}
    \caption{Phase space portrait of Eq.~\eqref{eq:ham12} (resonance $(1,2)$) with $G=J_2=1$, $\alpha_{12}=0$. The red line delimits the allowed circle, the blue line is the coupling arc.}
    \label{fig:phsp_12}
\end{figure}

Resonance $(1,2)$ Hamiltonian in $(\phi_1, J_1)$ coordinates, corresponding to the resonant Normal Form of a H\'enon-like map with a normal sextupolar kick, reads~\cite{PhysRevLett.110.094801,chao2015emittance},
\begin{equation}
\ham(\phi_1,J_1) = \delta J_1 + \alpha_{11}J_1^2 + \alpha_{12}J_1 J_2 + GJ_1^{1/2}(J_2-2J_1)\cos \phi_1\,. \label{eq:ham12}
\end{equation}

The phase space features an allowed circle given by $J_1<J_2/2$, and a coupling arc. From Eq.~\eqref{eq:ufp} one obtains the unstable fixed points as solutions of
\begin{equation}
    \cos\phi =\frac{\delta + (\alpha_{11} + \alpha_{12})J_2}{G\sqrt{2J_2}} 
\end{equation}
and a coupling arc (see Eq.~\eqref{eq:couplarc}) that, expressed in Cartesian coordinates, reads
\begin{equation}
    4\alpha_{11}(X^2+Y^2) - \frac{G}{\sqrt{2}}X + 2(\delta + \alpha_{12}J_2+2\alpha_{11}J_2)=0\,.
\end{equation}
This represents a circumference that crosses the allowed circle  when
\begin{equation}
    \qty|\frac{\delta + (\alpha_{11} + \alpha_{12})J_2}{G\sqrt{2J_2}}| \le 1
\end{equation}
dividing it in two regions. When $\alpha_{11}=0$, the coupling arc reduces to the straight line 
\begin{equation}
    X = \frac{\sqrt{2}(\delta + \alpha_{12}J_2)}{2G}
\end{equation}
that sweeps through the phase space if $\delta$ is varied, defining two equal regions when $\delta=-\alpha_{12} J_2$,. The equation of the stable fixed points for $\phi_1=0$ or $\phi_1=\pi$ reads
\begin{equation}
    (\delta + 2\alpha_{11}J_1 + \alpha_{12}J_2)J_1^{1/2} \pm G(J_2-6J_1)=0 \, , 
\end{equation}
and we obtain two real solutions inside the allowed circle, one for each side of the coupling arc. Therefore, the phase space is always divided into no more than two regions. Some phase space portraits are shown in Fig.~\ref{fig:phsp_12}.

\subsubsection{Resonance \texorpdfstring{$(2,1)$}{(2,1)}}

\begin{figure}
    \centering
    \includegraphics[width=.43\textwidth]{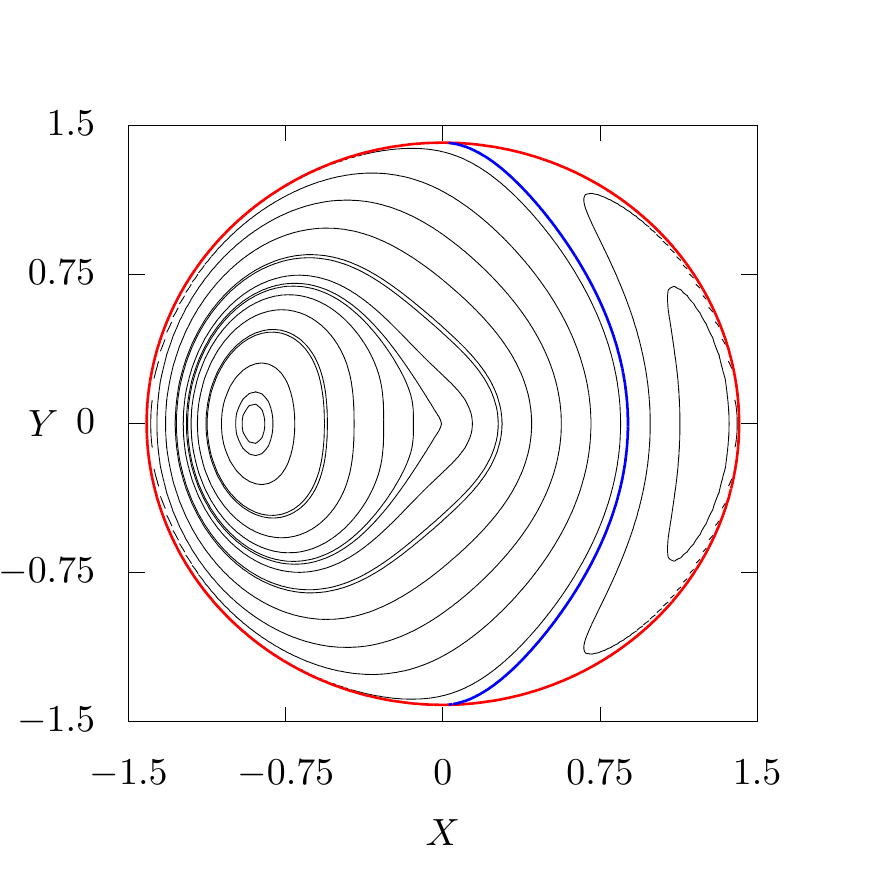}
    \caption{Phase space portrait of Eq.~\eqref{eq:ham21} (resonance $(2,1)$)  with $\delta=G=J_2=1$, $\alpha_{11}=\alpha_{12}=0$. The red line delimits the allowed circle, the blue line is the coupling arc.}
    \label{fig:phsp_21}
\end{figure}

The starting point is the H\'enon-like 4D map with a skew sextupole kick and the resonant Normal Form provides an interpolating Hamiltonian up to order 3 of the form
\begin{equation}
\ham(\phi_1,J_1) = \delta J_1 + \alpha_{11}J_1^2 + \alpha_{12}J_1J_2+ 2G J_1\sqrt{J_2-J_1}\cos\phi_1 \, ,
\label{eq:ham21}
\end{equation}
and the motion is limited to the allowed circle $J_1<J_2$. The fixed points on the allowed circle are given by $\cos\phi_1=0$, \ie $\phi_1=\pm\pi/2$, whereas the expression of the coupling arc is obtained by solving $\ham(J_1,\phi_1) - \ham(J_1=J_2,\phi_1=\pm\pi/2)$, \ie
\begin{equation}
 (\delta+\alpha_{11}(J_2+J_1)+\alpha_{12}J_2)\sqrt{J_2-J_1} = 2GJ_1\cos\phi_1\,,
\end{equation}
which is easily solved when $\alpha_{11}=0$:
\begin{equation}J_1(\phi_1) = -\frac{\hat\delta^{2} - \hat\delta \sqrt{16  G^{2} J_{2}
\cos\left(\phi_{1}\right)^{2} + \hat\delta^{2}} }{8  G^{2}
\cos\left(\phi_{1}\right)^{2}} \qquad \text{with} \qquad \hat\delta = \delta+\alpha_{12}J_2 \,. 
\label{eq:29}
\end{equation}

We remark that if $\hat\delta>0$ we must have $\cos\phi_1>0$, \ie the coupling arc lies in the positive domain of $X$, whereas for $\hat\delta<0$ in the negative one. Moreover, for $\hat\delta=0$ the coupling arc reduces to a line that evenly divides the allowed circle. On the other hand, we can look for solutions when $\phi_1=0$ and $\phi_1=\pi$, and when $\alpha_{11}=0$, Eq.~\eqref{eq:29} reads
\begin{equation}
\hat\delta\sqrt{J_2-J_1} \pm G(2J_2-3J_1) = 0\,.
\end{equation}
Assuming $G>0$, we need to impose conditions on the existence of the solutions before squaring: for $\phi_1=0$ and $\hat\delta>0$, the condition $2J_2/3<J_1<J_2$ holds, while for $\hat\delta<0$ we require $J_1<2J_2/3$. For $\phi_1=\pi$ the conditions are reversed. Finally, we obtain the solutions
\begin{equation}
J_1^\pm = \frac{2}{3}J_2 \pm \frac{\hat\delta}{18G^2} \qty(\sqrt{12G^2 J_2+\hat\delta^2}\mp \hat\delta)\,.
\end{equation}

No matter the sign, the quantity inside the brackets is always positive, which implies $J_1^+>2J_2/3$ if $\hat\delta>0$ and $J_1^+<2J_2/3$ if $\hat\delta<0$, and this solution is acceptable only for $\phi_1=0$. Conversely, $J_1^->2J_2/3$ if $\hat\delta<0$ and $J_1^-<2J_2/3$ if $\hat\delta>0$. This solution is only acceptable when $\phi_1=\pi$. Finally, we always have a solution in the positive $X$ semi-axis and one in the negative one, as long as the solution for $J_1$ inside the allowed circle, but, as  $J_1^+ \to J_2$ when $\hat\delta \to \infty$, and $J_1^- \to J_2$ as $\hat\delta\to-\infty$, this never occurs.

Let us study the trajectory of a point whose initial condition is at the origin. We have to solve the equation $\ham(\phi_1,J_1) = 0$, \ie
\begin{equation}
J_1\qty(\delta + 2G\sqrt{J_2-J_1}\cos\phi_1)=0\,,
\end{equation}
and we have the solutions $J_1=0$ and $\delta + 2G\sqrt{J_2-J_1}\cos\phi_1=0$. The latter can only be solved for $\cos\phi_1<0$ if $\delta>0$, and $\cos\phi_1>0$ if $\delta<0$. Therefore, there is only one trajectory passing through the origin: it does not alter the topology of the phase space introducing new islands (see Fig.~\ref{fig:phsp_21}), and the allowed circle is always divided into two regions, thus making the emittance sharing possible.

We remark that in Fig.~\ref{fig:phsp_21} and in general in the phase-space portraits of the Hamiltonian functions discussed in this paper, we used large values of $\delta$ and $J_2$, compared to those chosen for the numerical simulations that will be later discussed. This is justified by the fact that the Hamiltonian models depend on the unique parameter $\eta=\delta/(G\sqrt{J_2})$, for third order resonances, and $\eta=\delta/(GJ_2)$, for fourth-order ones (see Eq.~\ref{eq:eta}), hence it is perfectly justified to choose conditions with $\eta\sim 1$.

\subsubsection{Resonance \texorpdfstring{$(1,3)$}{(1,3)}}

\begin{figure}
\centering
\includegraphics[width=.55\textwidth]{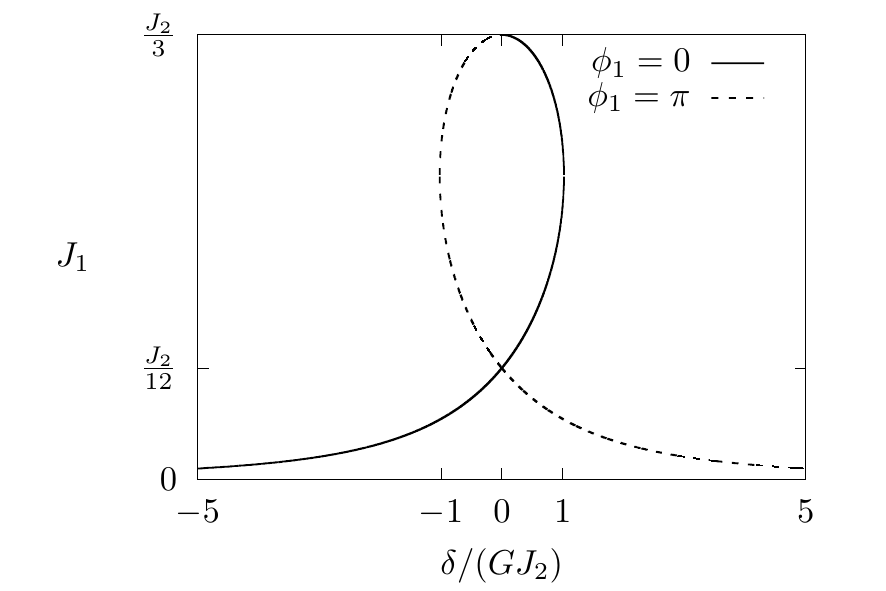}\\
\caption{Fixed points of Eq.~\eqref{eq:ham13} (resonance $(1,3)$) with  $\alpha_{11}=\alpha_{12}=0$ for $\phi_1=0$ (solid line) and $\phi_1=\pi$ (dashed line) as a function of $\delta$.}
\label{fig:fpr13}
\end{figure}

\begin{figure}[htb]
\centering
\includegraphics[trim=0truemm 0truemm 10truemm 0truemm,width=0.9\textwidth,clip=]{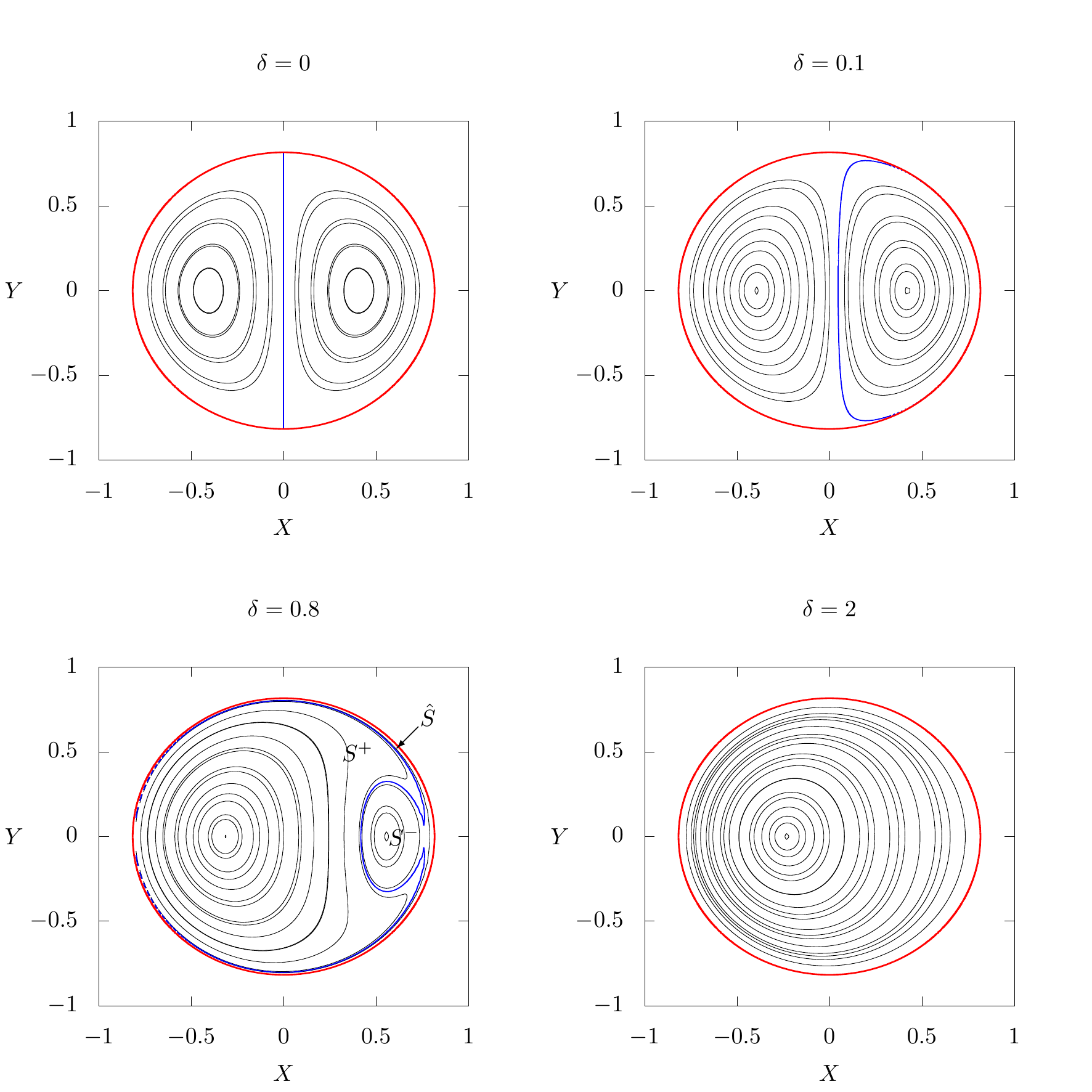}%

\caption{Phase space of Eq.~\eqref{eq:ham13} (resonance $(1,3)$) for different values of $\delta$ with $G=J_2=1$, $\alpha_{11}=\alpha_{12}=0$. The red line delimits the allowed circle, while the blue line is the separatrix. In the bottom-left plot, the extended region on the left is $S^+$, the one inside the separatrix on the right is $S^-$ while the small region between the separatrix and the allowed circle is $\hat S$.} 
\label{fig:phsp13}
\end{figure}

For the $(1,3)$ resonance, which is excited using a skew octupole, we have the quasi-resonant Hamiltonian
\begin{equation}
\ham(\phi_1,J_1) = \delta J_1 + \alpha_{12}J_1 J_2 + \alpha_{11}J_1^2 + GJ_1^{1/2}(J_2-3J_1)^{3/2}\cos\phi_1 \, .
\label{eq:ham13}
\end{equation}
If we set $\alpha_{11}=\alpha_{12}=0$, which is the case when the resonance is excited without sextupolar kicks, we have fixed points for $\phi_1=0$ or for $\phi_1=\pi$ that are the solutions of 
\begin{equation}
\pdv{\ham}{J_1}\eval_{\phi_1=0,\pi} = \delta \pm \frac{G}{2}\qty( J_1^{-1/2}(J_2-3J_1)^{3/2} - 9J_1^{1/2}(J_2-3J_1)^{1/2})=0 \, ,
\end{equation}
that gives
\begin{equation}
\delta J_1^{1/2} = \pm\frac{G}{2}\qty(9 J_1 (J_2-3J_1)^{1/2} - J_1^{1/2}(J_2-3J_1)^{3/2}) \, .
\label{eq:36}
\end{equation}
The \rhs of Eq.~\ref{eq:36} is positive when $\pm G\qty(J_1-\frac{J_2}{12})> 0$, and we will compare it to the sign of $\delta$. Let us choose $G>0$. For $\phi_1=0$, we have solutions for $\delta>0$ and $J_2/12<J_1<J_2/3$, or for $\delta<0$ and $0<J_1<J_2/12$. For $\phi=\pi$ the conditions are reversed. By squaring the equation, which gives a cubic polynomial, we compute its roots, taking into account all conditions. The solutions are given in Fig.~\ref{fig:fpr13}. There are the following possibilities:
\begin{itemize}
    \item if $\delta/(GJ_2)>1$, there exists only one stable fixed point for $\phi_1=\pi$ that tends to the origin when $\delta/(G J_2)\gg 1$;
    \item if $0< \delta/(GJ_2) < 1$, there are two fixed points on $\phi_1=0$ and one on $\phi_\pi$. The inner solution on $\phi_1=0$ ($J_1^+$) and the solution on $\phi_1=\pi$ ($J_1^-$) are stable, while the outer fixed point on $\phi_1=0$ is unstable, and generates a separatrix. The phase space is divided into three regions: $S^\pm$ around $J_1^\pm$, and $\hat S$ that is the area between the separatrix which crosses $\hat J_1$ and the allowed circle. Portraits with $\delta/(GJ_2)=0.1$ and $\delta/(GJ_2)=0.8$ are shown in Fig.~\ref{fig:phsp13};
    \item if $\delta=0$, two fixed points are present in $J_2/12$, at $\phi_1=0$ and $\phi_1=\pi$. The separatrix degenerates to the diameter of the allowed circle.
    \item if $\delta<0$, one has the same situation as for $\delta>0$, but exchanging $\phi_1=0$ and $\phi_1=\pi$.
\end{itemize}

\subsubsection{Resonance \texorpdfstring{$(3,1)$}{(3,1)}}

From the general properties stated before, the allowed circle is $J_1 <J_2$ and the coupling arc intersects the border of the allowed circle at $\phi_1=\pm\pi/2$. Then, we have the origin that, being $m>2$, is a stable fixed point.

For what concerns the fixed points on the $X$ axis, we initially consider the case with $\alpha_{11}=\alpha_{12}=0$. For $\phi_1=0$ or $\phi_1=\pi$, the equation $\pdv*{H}{J_1}=0$ reads
\begin{equation}
2\delta\sqrt{J_2-J_1}=\pm 3\sqrt{3}GJ_1^{1/2}(4J_1-3J_2)\,.
\end{equation}
Assuming $G>0$, for $\delta>0$ we can accept solutions on $\phi_1=0$ for $J_1<3J_2/4$ and on $\phi_1=\pi$ for $3J_2/3<J_1<J_2$, and the opposite for $\delta<0$. By squaring, we obtain the cubic equation
\begin{equation}
4\delta^2(J_2-J_1)=27 G^2 J_1(4J_1-3J_2)^2
\end{equation}
whose roots can be studied by rewriting the equation as
\begin{equation}
\frac{4\delta^2}{27G^2}=\frac{J_1(4J_1-3J_2)^2}{J_2-J_1}=f\qty(\frac{J_1}{J_2})\,,
\label{eq:r31d2}
\end{equation}
and by studying $f(J_1/J_2)$ as a function of $J_1$ in $[0,J_2]$. This function has zeroes in $J_1=0$ and $J_1=3J_2/4$ and diverges to $+\infty$ as $J_1\to J_2$. From its derivative, we find that a local maximum exists for $J_1=(3-\sqrt{3})J_2/4$ and the corresponding value of $\delta$ is (see Eq.~\eqref{eq:r31d2}) 
\begin{equation}
\delta^* =\pm \frac{9G}{2}\sqrt{2\sqrt{3}-3}J_2\approx \pm 3.1GJ_2\,.
\end{equation}

\begin{figure}
    \centering
    \includegraphics[height=7cm]{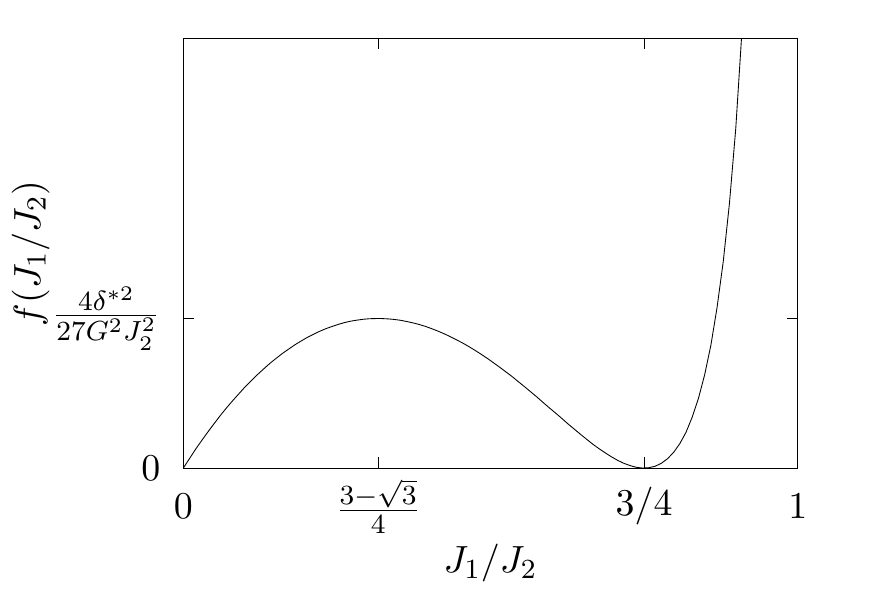}
  \caption{Plot of $f(J_1/J_2)= \dfrac{(J_1/J_2)\qty(4J_1/J_2-3)^2}{1-J_1/J_2}$, as introduced in Eq.~\eqref{eq:r31d2}. The real solutions of the equation are found as $f(J_1/J_2) = 4\delta^2/(27G^2J_2)$.}
    \label{fig:funr31}
\end{figure}

\begin{figure}
    \centering
    \includegraphics[trim=0truemm 0truemm 10truemm 0truemm, width=\textwidth]{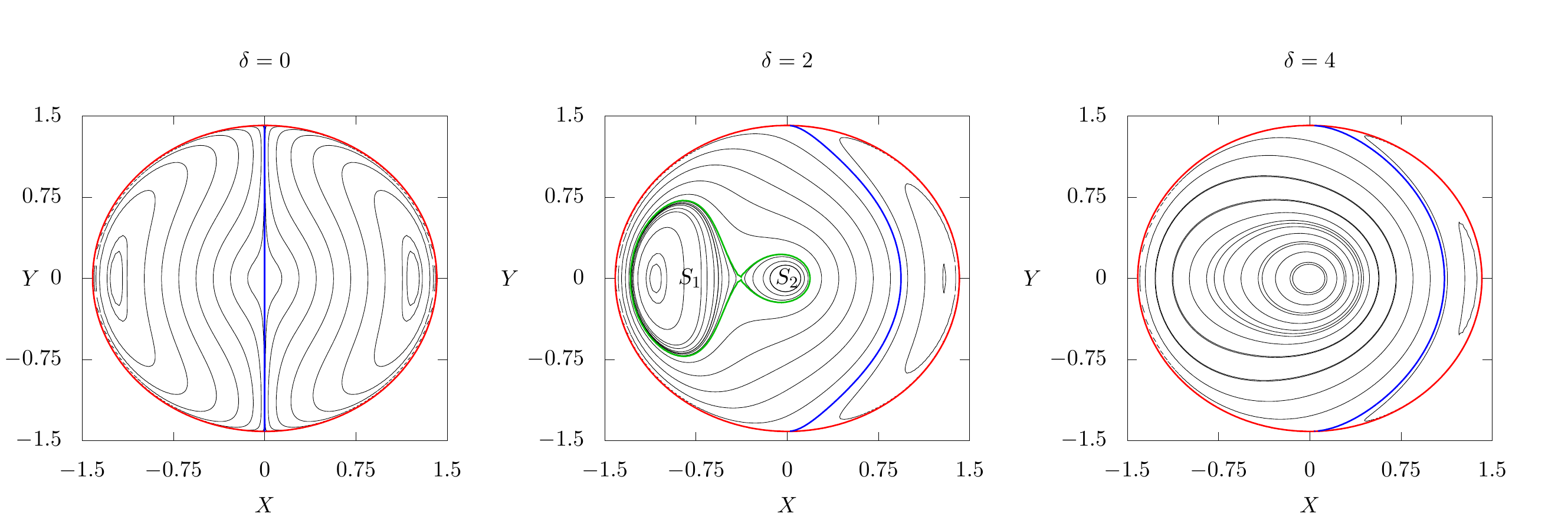}
    \caption{Phase space of Hamiltonian of Eq.~\eqref{eq:ham_mn_J1J2}, with $m=3$, $n=1$, expressed in ($X=\sqrt{2J_1}\cos\phi_1$, $Y=\sqrt{2J_1}\sin\phi_1$) coordinates, for three values of $\delta$, having set $G=J_2=1$ and the amplitude  detuning coefficients to zero. The red line delimits the allowed circle while the blue line is the coupling arc. In plot for $\delta=2$, the green line is the extra separatrix which delimits the regions $S_1$ and $S_2$.}
    \label{fig:ph31}
\end{figure}

The plot of $f(J_1/J_2)$ is shown in Fig.~\ref{fig:funr31}. Considering the sign conditions on the solution, one has the following possibilities (some examples of phase-space portraits are shown in Fig.~\ref{fig:ph31}):
\begin{itemize}
\item if $\delta>\delta^*$, there are a stable fixed point at the origin and a stable fixed point at the right of the coupling arc for $\phi_1=0$ and $J_1>3J_2/4$ (see Fig.~\ref{fig:ph31}, right);
\item if $0<\delta<\delta^*$, a stable fixed point at the origin, an  unstable fixed point for $\phi_1=\pi$, $0<J_1<(3-\sqrt{3})J_2/4$, and a stable fixed point for $\phi_1=\pi$ and $(3-\sqrt{3})J_2/4<J_1<3J_2/4$, plus a stable fixed point at the right of the coupling arc, for $\phi_1=0$ and $J_1>3J_2/4$. The separatrix that passes through the unstable fixed point is the green line in Fig.~\ref{fig:ph31} (centre) delimiting the regions $S_1$ and $S_2$;
\item if $\delta=0$, two stable fixed points at $J_1=3J_2/4$; the coupling arc is a line that passes through the origin (see Fig.~\ref{fig:ph31}, left);
\item if $-\delta^*<\delta<0$, a stable fixed point at the origin, an unstable fixed point for $\phi_1=0$, $0<J_1<(3-\sqrt{3})J_2/4$, and a stable fixed point for $\phi_1=0$ and $(3-\sqrt{3})J_2/4<J_1<3J_2/4$, plus a stable fixed point at the left of the coupling arc, for $\phi_1=\pi$ and $J_1>3J_2/4$. The topology is the same of Fig.~\ref{fig:ph31} (centre), but horizontally reversed;
\item $\delta<-\delta^*$: a stable fixed point at the origin and a stable fixed point at the left of the coupling arc for $\phi_1=\pi$ and $J_1>3J_2/4$. Once more, the topology is mirrored \wrt the rightmost plot of Fig.~\ref{fig:ph31}.
\end{itemize}

\subsection{Emittance-sharing process} \label{sec:sharingproc}

\subsubsection{General considerations}

Let us consider a process described by the Hamiltonian of Eq.~\eqref{eq:initial_ham}, with either $\omega_x$ or $\omega_y$, or both, slowly changing as a function of time to cross the $(m,\,n)$ resonance. According to the transformations that led to Eq.~\eqref{eq:ham_mn_J1J2}, this is modelled varying $\delta$ from a case where $\hat\delta=\delta+\alpha_{12}J_2\gg 0$ to one where $\hat \delta\ll 0$, \ie $\hat\delta$ is adiabatically changed from $+\delta_\text{max}$ to $-\delta_\text{max}$ during a time interval $T$. The variation of $\hat\delta$ changes the position of the coupling arc, that sweeps the disk $J_1<J_2/n$ within which the dynamics is constrained.

A particle starts its orbit far from the resonance, with an action $J_{1,\text{i}} = J_{x,\text{i}}/m$, where, the only fixed point is close to the origin and the particle orbit is almost a circle, of area $2\pi \Ji{1}$. This area is an adiabatic invariant, and it is conserved when $\hat\delta$ is slowly varied. As $\hat\delta$ is decreased, the moving coupling arc reduces the extent of the region the particle is orbiting in, dividing the allowed circle in two parts that have equal area when $\hat \delta=0$. When the area of the initial region is equal to $2\pi \Ji{1}$, according to separatrix crossing theory~\cite{neish1975}, the particle will cross the coupling arc and enter the other phase-space region with an action corresponding to the area of the arrival region at the jump time divided by $2\pi$.

Since the allowed circle has an area $2\pi J_2/n$, the resulting action will be
\begin{equation}
    J_{1,\text{f}} = \frac{J_2}{n}-J_{1,\text{i}}\,, \end{equation}
and, transforming back to the initial actions
\begin{equation} \Jf{x} = m \Jf{1} = m \qty(\frac{\Ji{y}+ n \Ji{x}/m}{n}-\frac{\Ji{x}}{m}) = \frac{m}{n}\Ji{y} 
\label{eq:Jfx}
\end{equation}
and
\begin{equation} \Jf{y} = \frac{n}{m} \Ji{x}\,.
\label{eq:Jfy}
\end{equation}

As $\delta$ continues decreasing, the area of the region containing the particle orbit increases and, at the end of the variation of $\delta$ (far from the resonance), the orbit is a circle around the origin whose area corresponds to the new action.

For each particle, this process realises an \textit{action sharing} between the two degrees of freedom. The product $J_x J_y$ remains constant, but the two values are, at the end of the process, reallocated according to a $n/m$ ratio. Note that for the case of the linear coupling resonance, \ie $n=m=1$, this corresponds to the well-known emittance exchange process~\cite{Metral:529690,PhysRevAccelBeams.23.044003,PhysRevAccelBeams.24.094002}. It is essential to stress that the analysis outlined before holds true only when the phase space is exactly divided into two regions by the coupling arc, and no other separatrices are present. Otherwise, a different analysis is needed to assess whether the additional phase-space regions, such as the ones visible in the centre plot of Fig.~\ref{fig:ph31}, interfere with the trapping process leading to the emittance sharing. A discussion on this and how to mitigate such effects is carried out in Section~\ref{sec:topology_sharing}. If the action sharing is successful, it is possible to verify what happens in the presence of a set of initial conditions that are Gaussian distributed in both planes $(x, p_x)$ and $(y,p_y)$, \ie an exponential distribution in $J_x$ and $J_y$. Using the standard definition, \ie  $\eps_x = \av{J_x}$, $\eps_y = \av{J_y}$, the initial distribution reads
\begin{equation}
    \rho_\text{i}(J_x,J_y)  = \frac{1}{\eps_{x}\eps_{y}}\exp( -\frac{J_x}{\eps_{x}} - \frac{J_y}{\eps_{y}})
\end{equation}
and, after the exchange process using Eqs.~(\ref{eq:Jfx},~\ref{eq:Jfy}), we obtain the final distribution
\begin{equation}
    \rho_\text{f}(J_x,J_y)  = \frac{1}{\eps_{x}\eps_{y}}\exp( -\frac{m}{n}\frac{J_y}{\eps_{x}} - \frac{n}{m}\frac{J_x}{\eps_{y}}) \, .
\end{equation}
The new averages are given by the integrals
\begin{equation}
\begin{aligned}
     \eps_{x,\text{f}} = \av{\Jf{x}} & = \int_0^\infty\dd J_x\int_0^\infty \dd J_y\, J_x\, \rho_\text{f}(J_x,J_y)  & = \frac{m}{n}\av{\Ji{y}} = \frac{m}{n}\eps_{y,\text{i}} \\
     \eps_{y,\text{f}} = \av{\Jf{y}} &= \int_0^\infty\dd J_x\int_0^\infty \dd J_y\, J_y\, \rho_\text{f}(J_x,J_y)  & =\frac{n}{m}\av{\Ji{x}} = \frac{n}{m}\eps_{x,\text{i}} \, ,
\end{aligned}
\label{eq:emshar}
\end{equation}
and it is evident that an {\sl emittance sharing} occurred.

It is also possible to compute the initial distributions in terms of $J_1$ and $J_2$
\begin{equation}
\begin{aligned}
    \rho_1(J_1) & =\int_0^\infty \dd J_y \, \rho_\text{i}(mJ_1,J_y) = \frac{1}{\eps_x}
\exp(-\frac{mJ_1}{\eps_x}) \\
    \rho_2(J_2) & = \int_0^{\frac{m}{n}J_2}\dd J_x\, \rho_\text{i}\qty(J_x,J_2-\frac{n}{m}J_x) = \frac{m}{m\,\eps_x-n\,\eps_y}\qty[\exp(-\frac{J_2}{\eps_y}) - \exp(-\frac{m\,J_2}{n\,\eps_x})]\, .
\end{aligned}
\end{equation}
%
%and
%
%\begin{equation}
%    \rho_2(J_2) = \int_0^{\frac{m}{n}J_2}\dd J_x\, \rho_\text{i}\qty(J_x,J_2-\frac{n}{m}J_x) = \frac{m}{m\eps_x-n\eps_y}\qty[\exp(-\frac{J_2}{\eps_y}) - \exp(-\frac{mJ_2}{n\eps_x})]\, .
%\end{equation}
%
Then, given the dependence of the phase-space topology on the conserved parameter $J_2$, it is useful to consider the initial Gaussian distribution in $J_x$ and $J_y$ as an ensemble of distributions in $J_1$ dependent on the parameter $J_2$ distributed as $\rho_2(J_2)$: the distribution of $J_1$ for a given $J_2$ reads
\begin{equation}
    \rho_{12}(J_1|J_2) = \frac{\rho(mJ_1, J_2-nJ_x/m)}{\rho_2(J_2)} = \frac{m\,\eps_y - n\,\eps_x}{m\,\eps_x \eps_y} \frac{\exp(\frac{n-m}{\eps_x}J_1)}{1 - \exp(\frac{n\,\eps_x-m\,\eps_y}{n\,\eps_x\eps_y}J_2)}
\end{equation}
where the normalisation
\begin{equation}
\int_0^\infty \dd J_2\, \rho_2(J_2) \int_0^{J_2/n} \dd J_1 \, \rho_{12}(J_1|J_2) = 1
\end{equation}
holds.

%It is worth stressing that, depending on the initial conditions, each particles has its own quasi-conserved value of $J_2$, and, since the phase space topology depends on $J_2$, all the particles with the same $J_2$ will ``observe'' the same phase space, and, on the other hand, particles with different values of $J_2$ could have different behaviours, as, from dimensional analysis on Hamiltonian~\eqref{eq:ham_mn_J1J2}, one can define an adimensional parameter 
%\begin{equation}
%\eta = \frac{\delta}{G J_2^{\frac{m+n-2}{2}}}
%\end{equation}
%which determines the interplay between $\delta$, $G$ and $J_2$. Therefore, 

During the emittance-sharing process, $\delta$ is varied between $\pm \delta_\text{max}$, and correspondingly, $\eta$ (see Eq.~\eqref{eq:eta}) varies between $\pm \eta_\text{max}$, where $\eta_\text{max}=\eta(\delta_\text{max})$. For any pair $(J_{1,\text{i}},J_{2,\text{i}})$, there exists a value $\eta^*$ for which the area of the phase-space region $A_{\Ji_{2}}(\eta)$ satisfies $2\pi \Ji{1}=A_{\Ji{2}}(\eta^*)$, and whenever the phase space is divided into two regions, $A_{\Ji{2}}(\eta)$ is a monotonic decreasing function of $\eta$ (and of $\delta$) during the resonance-crossing process. Therefore, the function $J_1(\eta^*)=A(\eta^*)/(2\pi)$ is monotonic as well. During the resonance crossing, the fraction $\tau$ of particles that effectively undergoes emittance sharing is given by all particles for which $\eta^*\in [-\eta_\text{max},\eta_\text{max}]$ and it can be obtained by
\begin{equation}
    \tau = \int_0^\infty \dd J_2 \,\rho_2(J_2) \int_{J_1(-\eta_\text{max})}^{J_1(\eta_\text{max})}\dd J_1\,\rho_{12}(J_1) \, .
\end{equation}
The sharing fraction $\tau$ will also be a monotonic function of $\eta_\text{max}$. The parameter $\eta_\text{max}$ determines the effectiveness of the emittance sharing due to geometrical reasons: under the assumption that the initial beam distributions are Gaussian, one can define the following parameter
\begin{equation}
\kappa_\text{geom} = \frac{\delta_\text{max}}{G \av{\Ji{2}}^{(m+n-2)/2}} \,.
\label{eq:kgeom}
\end{equation}
%
%if $\av{\Ji{x}}=\av{\Ji{y}}=\eps_0$,  of course, $\av{\Ji{2}}\propto \eps_0$.
as the relevant quantity to study the performance of the emittance-sharing process.

The phase-space geometry is certainly important in the emittance-sharing process, but the efficiency is also influenced by the adiabaticity of the resonance-crossing process. A form for the adiabaticity parameter should therefore be determined. For this purpose we remark that the Hamiltonian of Eq.~\eqref{eq:ham_mn_J1J2} can be written, while $\delta$ is varied, as 
\begin{equation}
    \ham = \epsilon t J_1 + H_0(J_1) + G H_1(\phi_1,J_1)\,,
\end{equation}
where $\epsilon=2\delta_\text{max}/T$, and $H_0, H_1$ represent the amplitude-dependent and resonant terms, respectively, that appear in the equations of motion
\begin{equation}
\begin{split}
    \dot J_1    & = -G \, \pdv{H_1}{\phi} \\
    \dot \phi_1 & = \epsilon t + \pdv{H_0}{J} + G \, \pdv{H_1}{J} \,.
\end{split}
\end{equation}
As shown in Ref.~\cite{PhysRevAccelBeams.24.094002}, under the rescaling of time $\bar t = Gt$, one obtains the equations
\begin{equation}
\begin{split}
    \pdv{J_1}{\bar t}    & = -\pdv{H_1}{\phi} \\
    \pdv{\phi_1}{\bar t} & = \frac{\epsilon}{G^2} \bar t + \frac{1}{G}\pdv{H_0}{J} + \pdv{H_1}{J} \,.
\end{split}
    \label{eq:gscaling}
\end{equation}
Therefore, the appropriate adiabaticity parameter is given by $\epsilon/G^2$, \ie one obtains the same emittance sharing if $G$ scales as $G\sim \sqrt{\epsilon}$, while the amplitude-detuning terms are rescaled by a factor $G$. Parenthetically, as discussed in Ref.~\cite{PhysRevAccelBeams.24.094002}, it is possible to improve the adiabaticity of the resonance-crossing process by using $\delta \sim (\epsilon t)^p $ with $p>1$. If $\epsilon$ is kept constant and the sharing efficiency is evaluated for different values of $\delta_\text{max}$, then the parameter that controls the emittance sharing, including the dynamical effects, is given by 
\begin{equation}
\kappa_\text{dyn} = \frac{\sqrt{\delta_\text{max}}}{G \av{\Ji{2}}^{(m+n-2)/2}}\, .
\label{eq:kdyn}
\end{equation}
We remark that $\kappa_\text{geom}/\kappa_\text{dyn} = \sqrt{\delta_\text{max}}$.

Note that an effective resonance strength, which corresponds to the inverse of the parameter $\kappa_\text{dyn}$ defined above, was introduced in Ref.~\cite{PhysRevLett.110.094801}~and~\cite{chao2015emittance} as the unique parameter needed to describe the emittance sharing due to the crossing of the resonance $(1, 2)$. Our discussion shows that the purely phenomenological choice can be explained by means of rigorous mathematical arguments.

%In the following we proceed to analyse some resonances that can be excited using magnetic elements commonly installed in particle accelerators. Note that a discussion about which magnet excites a given resonance, carried out using Normal Form theory, is found in Appendix~\ref{app:magnets}.

\subsubsection{Effect of phase-space topology on emittance sharing} \label{sec:topology_sharing}

A general assumption on emittance sharing requires that the allowed circle is divided by the coupling arc in two regions. From the considerations reported in Section~\ref{sec:resonances}, this is always true for third-order resonances. However, for fourth-order resonances, such as $(1,3)$ and $(3,1)$, the situation is more complex. Indeed, close to the resonance $(1,3)$, an extra phase-space region is present (see Fig.~\ref{fig:phsp13}), although it does not affect the emittance sharing. 

Let us follow the evolution of the system from a state when $\delta \gg GJ_2$ and one with $\delta\ll -GJ_2$. At the beginning, only a fixed point is present, around which the particle orbits. When $\delta<GJ_2$, the region $\hat S$ appears (see Fig.~\ref{fig:phsp13}) and particles orbiting outside the new separatrix are automatically captured into that region, without any jump in $J_1$, since the area they enclose within their orbit remains the same.

While $\delta$ further decreases, however, $\hat S$ is pushed towards the outer circle. Particles inside it are then captured into $S^+$, for which $\Theta^+=\dv*{A(S^+)}{\delta}>0$, with the expected change in the adiabatic invariant. However, since in the crossing of the outer separatrix no change of adiabatic invariant occurs, the passage from $\hat S$ to $S^-$ is perfectly equivalent to the passage between $S^+$ and $S^-$. Once $\delta$ reaches zero, the situation is perfectly symmetric, with two stable fixed points and a separatrix dividing the allowed circle in two equal parts. 

We then continue reducing $\delta$ in the negative domain. A new unstable fixed point appears at $\phi_1=\pi$, and a topology akin to the third plot of Fig.~\ref{fig:phsp13}, although mirrored, appears. The problem is whether the new outer region will trap particles, and this turns out not to be possible. The outer region is maximal at $\delta=-G J_2$, and the unstable fixed point is at $J_1=J_2/4$ and $\phi_1=\pi$. We can thus estimate the area of the outer region as the difference between the outer circle at $J_1=J_2/3$, and the circle at $J_1=J_2/4$, which gives $\pi J_2/6$. On the contrary, particles inside $\hat S$ have a minimum action of $J_2/4$, \ie their orbit area is at least $\pi J_2/2$. Hence, since the area of $\hat S$ is always smaller than $\pi J_2/2$, no particle can reach the minimum action required when crossing from $S^+$ to $\hat S$. Thus, $\hat S$ remains void until, at $\delta=-GJ_2$ it disappears completely. 

Finally, the extra fixed point does not affect the emittance exchange process, as all particles pass from $S^-$ to $S^+$, which results, according to our previous generic analysis, in an emittance exchange.

In the case of the resonance $(3,1)$, the presence of extra stable fixed points (see Fig.~\ref{fig:ph31}) translates in an extra (and unavoidable) phase-space region that can, in principle, trap particles, thus spoiling the emittance sharing. Nevertheless, numerical observations discussed in Section~\ref{sec:num_res}) show that emittance sharing is still feasible, although with some reduction in performance due to the particles trapped in the extra region.

\section{Results of numerical simulations} \label{sec:num_res}

To assess the performance of the emittance-sharing process for different resonances, we compute the evolution of a Gaussian distribution of initial conditions $\rho(J_x,J_y)$ under the dynamics generated by the map of Eq.~\eqref{eq:henon} iterated for a number $N$ of turns, with or without amplitude-detuning terms, where $\omega_x$ is kept constant while $\omega_y$ is linearly varied between an initial value $\omega_{y,\text{f}} = m\omega_y/n + \delta_\text{max}$ and the final one $\omega_{y,\text{f}} = m\omega_y/n - \delta_\text{max}$, to cross the $(m,\, n)$ resonance. The initial and final emittance values are then compared, and a figure of merit is introduced to evaluate how well the emittance sharing occurred. To do so, we adapt the definition of $P_\text{na}$, introduced in Ref.~\cite{PhysRevAccelBeams.23.044003} for the $(1,\, 1)$ resonance crossing leading to emittance exchange  when a $(m,\, n)$ resonance is crossed, and emittance sharing occurs. The generalised definition is 
\begin{equation}
    P_\text{na} = 1 -  \frac{ \av{\Jf{x}} - \av{\Ji{x}}}{(m/n)\av{\Ji{y}} - \av{\Ji{x}}} \, .
    \label{eq:pna}
\end{equation}
The value of $P_\text{na}$ is $1$ when no emittance sharing is achieved and zero when the sharing is perfect, \ie according to the Eq.~\eqref{eq:emshar}.

Note that in Ref.~\cite{PhysRevLett.110.094801} a different figure of merit is introduced, the so-called fractional emittance growth (FEG), defined as
\begin{equation}
    \mathrm{FEG} = \left| \frac{\av{\Jf{x}}}{\av{\Ji{x}}}-1\right|+\left|\frac{\av{\Jf{y}}}{\av{\Ji{y}}}-1\right| \, , 
\end{equation}
which is $0$ when no exchange is performed and $\displaystyle{\left|\frac{m\av{\Ji{y}}}{n\av{\Ji{x}}} - \frac{n\av{\Ji{x}}}{m\av{\Ji{y}}}\right|}$ for a perfect emittance sharing. The reason of this definition is that the goal of~\cite{PhysRevLett.110.094801} is to avoid emittance sharing, and therefore to minimise the FEG, whereas our goal is the opposite, as we are looking to perform emittance sharing, and hence $P_\text{na}$ is the best choice for our study.

\begin{figure}[p]
    \centering
    \includegraphics[height=.9\textheight]{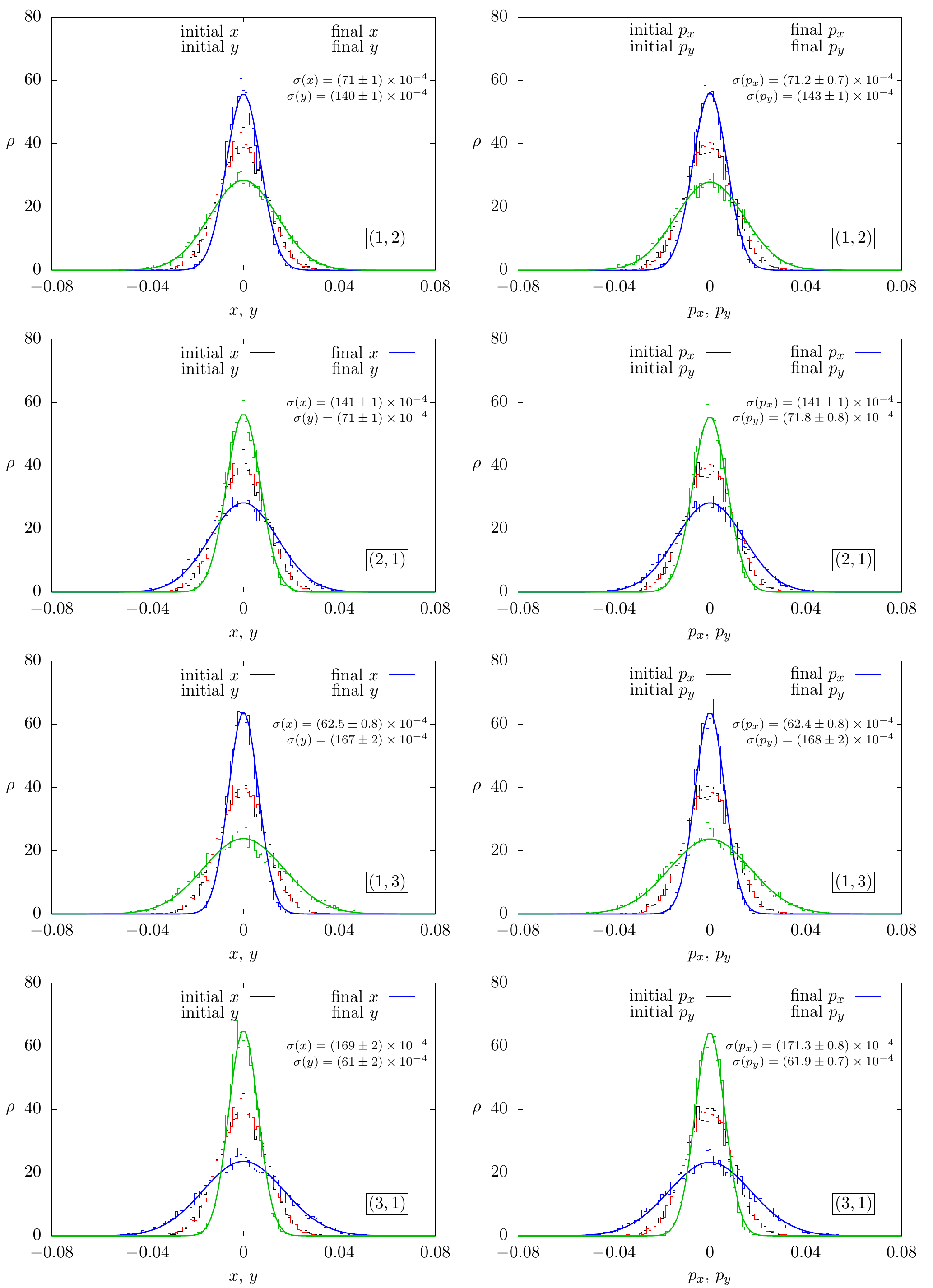}
    \caption{Histograms of the initial and final distribution of $x$, $y$ (left plots) and $p_x$, $p_y$ (right plots) after the resonance-crossing process for the four resonances under study. The initial distribution is Gaussian, with a standard deviation $\av{\Ji{x}}=\av{\Ji{y}}=0.01$. The map of Eq.~\eqref{eq:henon} has been used, with parameters $\delta_\text{max}=0.1$, $j_2=1$ or $k_2=1$, $N=10^6$, $\alpha=0$ for the third-order resonances; $\delta_\text{max}=0.01$, $j_3=10$, $N=10^7$, $\alpha=0$ for the fourth-order ones. The thick blue and green lines represent the Gaussian fits with zero average of the final distributions. The values of $\sigma$ printed on the plots represent the standard deviation of the Gaussian fits of the final distributions (for the initial distributions, $\sigma(x)=\sigma(y)=\sigma(p_x)=\sigma(p_y) = \num{100(1)e-4}$).}% The values of the fit parameters are shown in Table~\ref{tab:fit}.}
    \label{fig:distr}
\end{figure}

When not stated differently, we set $\beta_x=\beta_y=1$ and $\av{\Ji{x}}=\av{\Ji{y}}=\num{1e-4}$ and $\alpha=0$, generating distributions of initial conditions with $N_p=\num{1e4}$ particles. For the $(1,2)$ and $(2,1)$ resonances we use $\delta_\text{max}=0.1$, $k_2$ (or $j_2)$ equal to $1$, $N=\num{1e6}$. For the $(1,3)$ and $(3,1)$ resonances we set $\delta_\text{max}=0.01$, $j_3=10$, $N=\num{1e7}$. The initial and final distributions of $x$, $y$, $p_x$, $p_y$ for the four resonances are plotted in Fig.~\ref{fig:distr}, and they show clearly the effectiveness of the emittance sharing using these default parameters. Fits of Gaussian distributions with zero average are performed for the final distributions of the phase-space variables. The excellent agreement between the numerical results and the fitted functions shows that the emittance-sharing process preserves the Gaussian nature of the beam, acting only on the standard deviation.
% (the values of the fitted $\sigma$, assuming a Gaussian model, are shown in Table~\ref{tab:fit}).

%
%\begin{table}[b]
%    \centering
%    \caption{Gaussian fit parameters (standard deviations) for the final distributions of Fig.~\ref{fig:distr}, together with the value of $\av{\Jf{x}}$ and $P_\text{na}$ in each configuration.}
%    \begin{tabular}{|c|c|c|c|c|c|c|}
%    \hline
%    $(m,n)$ & $10^4\,\sigma(x)$ & $10^4\,\sigma(p_x)$ & $10^4\,\sigma(y)$ & $10^4\,\sigma(p_y)$ & $10^4\,\av{\Jf{x}}$ & $10^{-2}\,P_\text{na}$\\
%    \hline
%    $(1,2)$ & \num{71.4\pm	1.0} &	\num{71.2\pm 0.7} &	\num{140.1\pm 1.0} & \num{143.0\pm	1.4} & \num{0.505\pm 0.005} & \num{0.9\pm 1.0} \\
%    $(2,1)$ & \num{141.3\pm	1.0} &	\num{141.4\pm 1.3} &	\num{70.7\pm 1.0} & \num{71.8\pm	0.8} & \num{1.99\pm 0.02} & \num{0.5\pm 2.0} \\
%    $(1,3)$ & \num{62.5\pm	0.8} &	\num{62.4\pm 0.8} &	\num{167\pm 2} & \num{168.0\pm	1.7} & \num{0.378\pm 0.004} & \num{6.7\pm 0.6} \\
%    $(3,1)$ & \num{169.2\pm	1.7} &	\num{171.3\pm 0.8} &	\num{61.2\pm 1.6} & \num{61.9\pm 0.7} & \num{2.90\pm 0.03} & \num{4.9\pm 1.4} \\
%    \hline
%    \end{tabular}
%    \label{tab:fit}
%\end{table}
%

We proceed with the quantitative evaluation of the performance of the proposed technique by analysing how $P_\text{na}$ changes as a function of the parameters. In particular, we have concentrated our analyses on the dependence of $P_\text{na}$ on: \textit{(i)} the excursion of $\omega_y$, \ie $\delta_\text{max}$; \textit{(ii)} the strength of the non-linear magnets $j_2$, $k_2$, and $j_3$, depending on the resonance crossed; \textit{(iii)} the number of map iterations (turns) $N$; \textit{(iv)} the detuning parameter $\alpha$ (that has been chosen as $\alpha=\alpha_{xx}=\alpha_{yy}=-2\alpha_{xy}$ to mimic the amplitude detuning generated by normal octupoles as done in Ref.~\cite{PhysRevAccelBeams.24.094002}); \textit{(v)} the initial values of $\av{J_x}$ and $\av{J_y}$; \textit{(vi)} the ratio between $\av{\Ji{y}}$ and $\av{\Ji{x}}$. It is worth stressing that in our numerical investigation of the 4D parameter space, the dependence of $P_\text{na}$ is probed by changing one parameter at a time while keeping the others set to their nominal values. 

\begin{figure}[htb]
    \centering
    \includegraphics[trim=5truemm 0truemm 5truemm 0truemm, width=.43\textwidth]{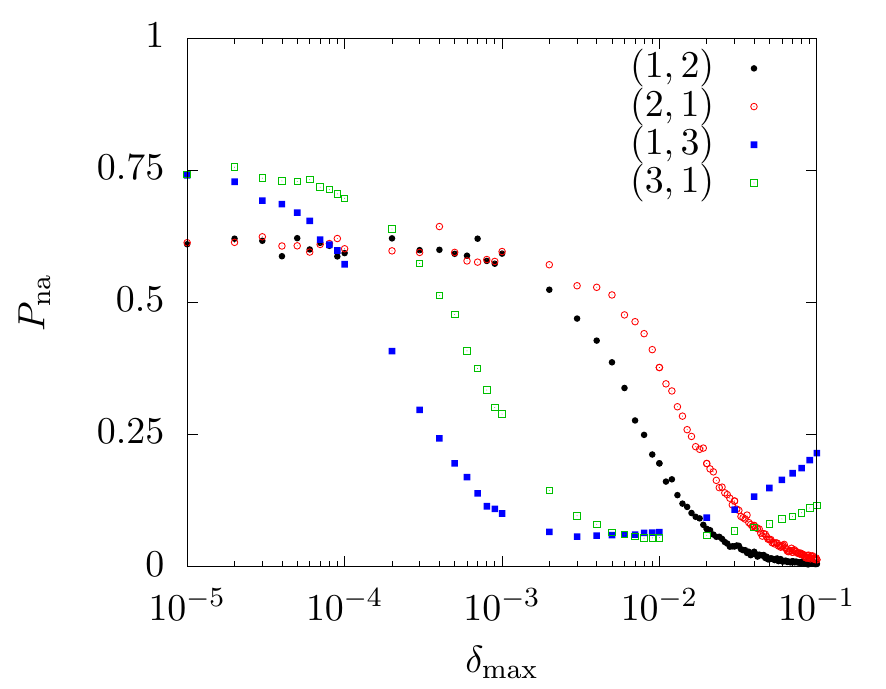} 
    \includegraphics[trim=0truemm 0truemm 5truemm 0truemm, width=.51\textwidth]{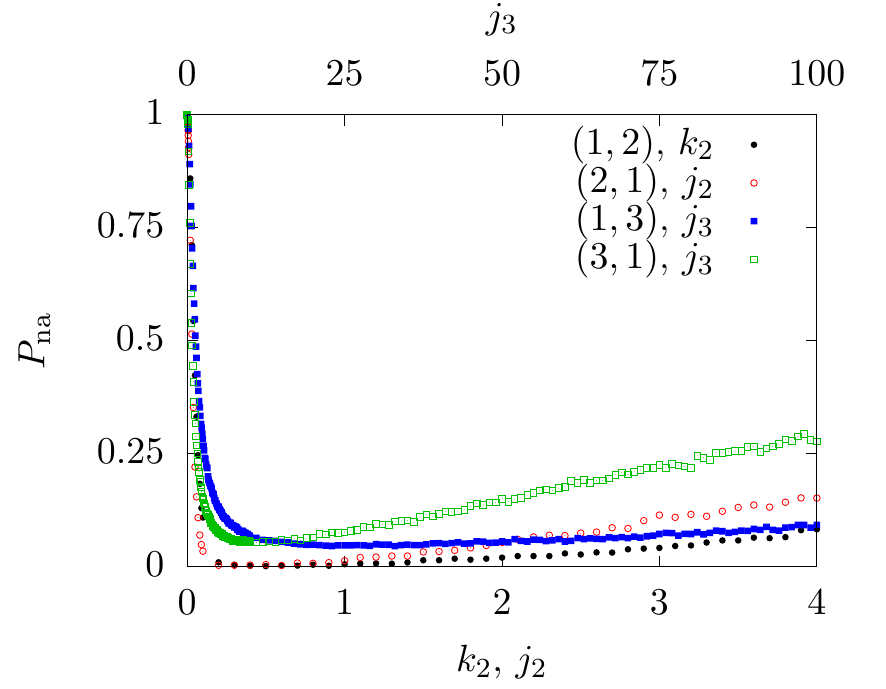}\\ 
     \caption{Left: $P_\text{na}$ as a function of the excursion $\omega_y$, \ie $\delta_\text{max}$. Right: $P_\text{na}$ as a function of the strength of the non-linear magnets ($k_2$, $j_2$ or $j_3$ depending on the resonance used). In both plots the map~\eqref{eq:henon} has been used, setting $\alpha=0$, and $N=10^6$ (for the third-order resonances) and $N=10^7$ (for the fourth-order ones), and using initial distributions with $\av{\Ji{x}}=\av{\Ji{y}}=0.01$. In the left plot, depending on the resonance, we set either $k_2=1$, $j_2=1$, or $j_3=10$. In the right plot, $\delta_\text{max}=0.1$ has been chosen for $(1,\, 2)$ and $(2,\, 1)$ while $\delta_\text{max}=0.01$ for $(1,\, 3)$ and $(3,\, 1)$.}
    \label{fig:plot_deltaG}
\end{figure}

Figure~\ref{fig:plot_deltaG} (left) shows the plot of $P_\text{na}$ as a function of $\delta_\text{max}$ whereas in the right part the dependence on the strength of the non-linear magnets is reported for the various resonances considered. A difference between the behaviour of the third- and fourth-order resonances is clearly seen. While for the former it is possible to determine the optimal value of $\delta_\text{max}$, or the strength of the non-linear magnets, such that $P_\text{na}=0$, \ie the emittance sharing is perfect, this is not the case for the latter resonances, for which $P_\text{na}$ reaches a non-zero minimum value.

We remark that when $\delta_\text{max}$ approaches $0$, $P_\text{na}$ does not converge to $1$. This is due to the fact that, when $\delta\approx 0$, the motion of all particles is very close to resonant conditions, and all particles revolve around one of the two stable fixed points in the phase space. Taking the average of the coordinate $J_1$ along the orbit, allows estimating the value of $P_\text{na}$ when $\delta_\text{max}\to 0$ (the details are found in Appendix~\ref{app:rescond}).

Other effects need to be taken into account, \eg the adiabaticity of the system. Weak non-linear coupling, which corresponds to a small value of $G$ in the Hamiltonian of Eq.~\eqref{eq:ham_mn_J1J2}, means a faster resonance crossing. For instance, for a particle moving close to the $(1,\, 2)$ resonance, the coupling arc,  given by the line of equation $X=\delta/(\sqrt{2}G)$, moves, over one map iteration, by $\delta X = \delta_\text{max}/(\sqrt{2}G N)$. This means that the adiabaticity condition is not met when the strength of the non-linear magnets is small, and for this reason $P_\text{na}$ goes to $1$. The same effect accounts for the lower sharing efficiency at large $\delta_\text{max}$, when $N$ is kept constant.

We remark that when the strength of the non-linear magnets becomes large, the quasi-resonant Hamiltonian may be no longer a good approximation of the dynamics generated by the map as the higher-order terms cannot be neglected anymore. This observation will be particularly relevant when we will discuss the results shown in Figs.~\ref{fig:combine_plot} later in this Section.

\begin{figure}[htb]
    \centering
    \includegraphics[trim=5truemm 0truemm 5truemm 0truemm, width=.43\textwidth]{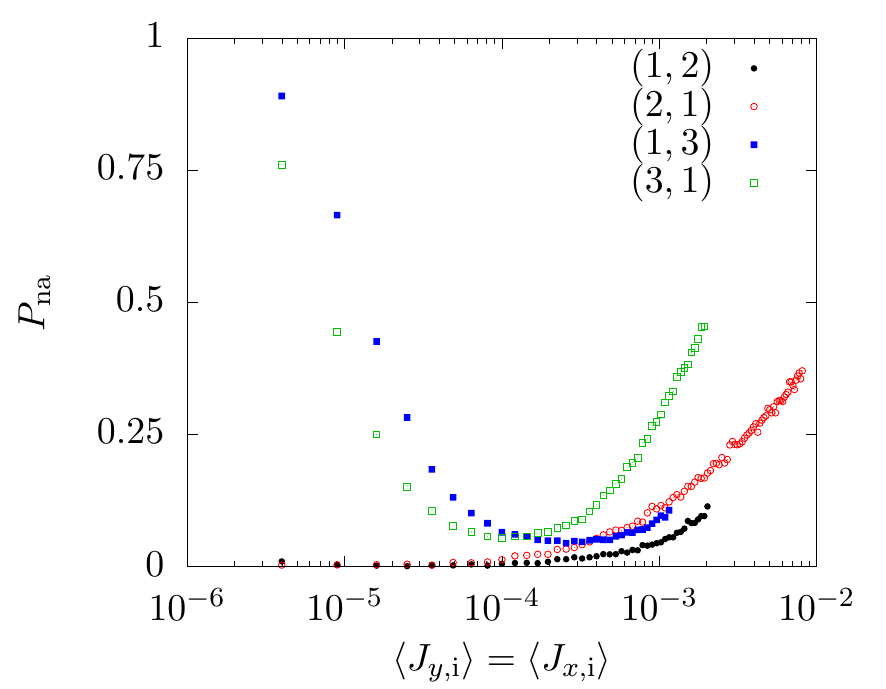} 
    \hspace{12truemm}
    \includegraphics[trim=5truemm 0truemm 5truemm 0truemm, width=.43\textwidth]{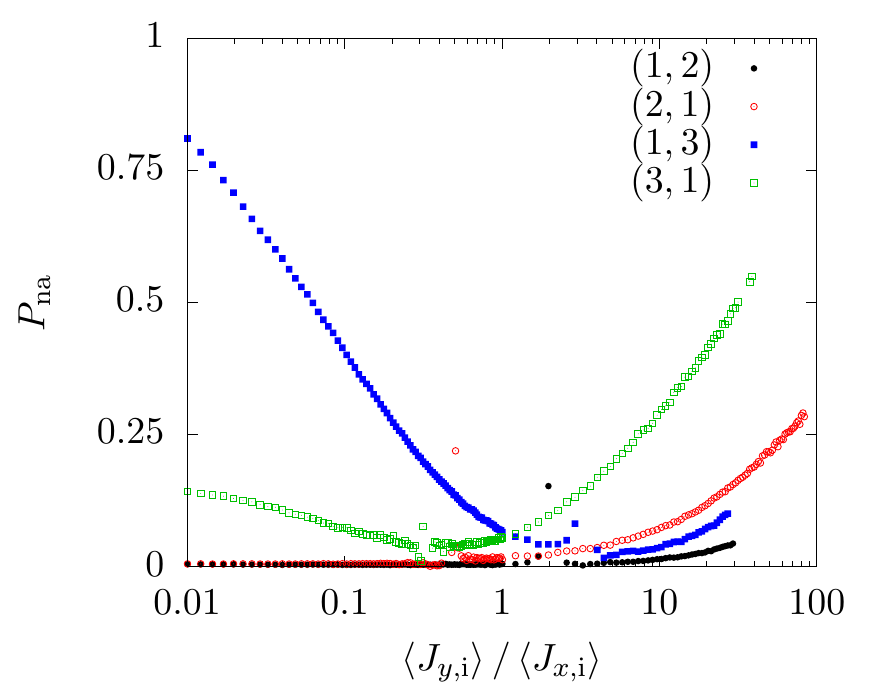}\\
     \caption{Left: $P_\text{na}$ as a function of the initial $\av{\Ji{x}}$ (chosen to be equal to $\av{\Ji{y}}$), for the four resonances. Right: $P_\text{na}$ as a function of the ratio between $\av{\Ji{y}}$ and $\av{\Ji{y}}$, using $\av{\Ji{x}}=0.01$. In both plots the map~\eqref{eq:henon} has been used, setting $\alpha=0$, and $\delta_\text{max}=0.1$, $j_2=1$ or $k_2=1$, $N=10^6$ (for the third-order resonances) and $\delta_\text{max}=0.01$, $j_3=10$, $N=10^7$ (for the fourth-order ones).}
    \label{fig:plot_sigma}
\end{figure}

\begin{figure}[htb]
    \centering
    \includegraphics[trim=5truemm 0truemm 5truemm 0truemm, width=.45\textwidth]{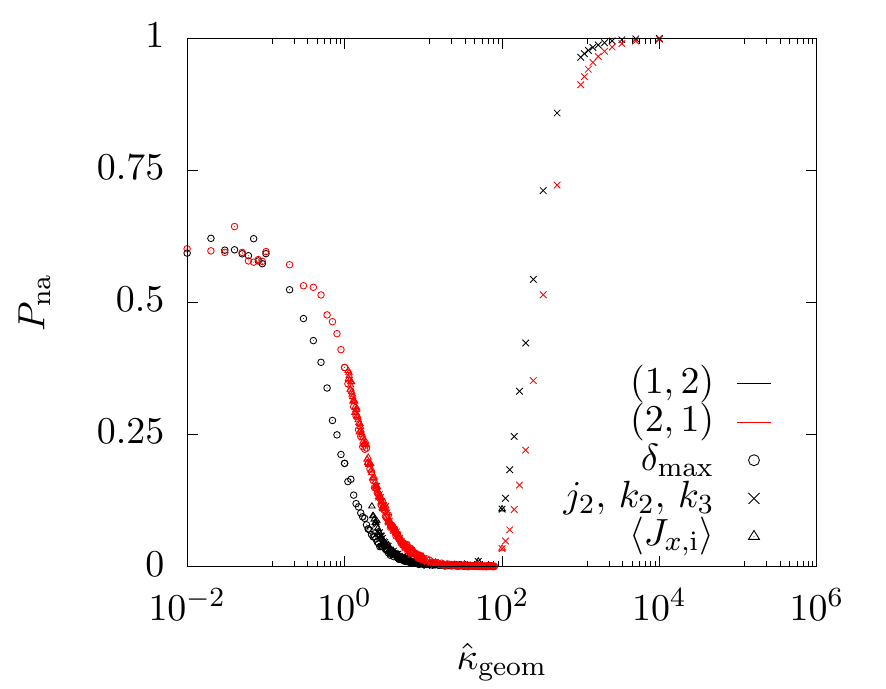}
    \hspace{10truemm}
    \includegraphics[trim=5truemm 0truemm 5truemm 0truemm, width=.45\textwidth]{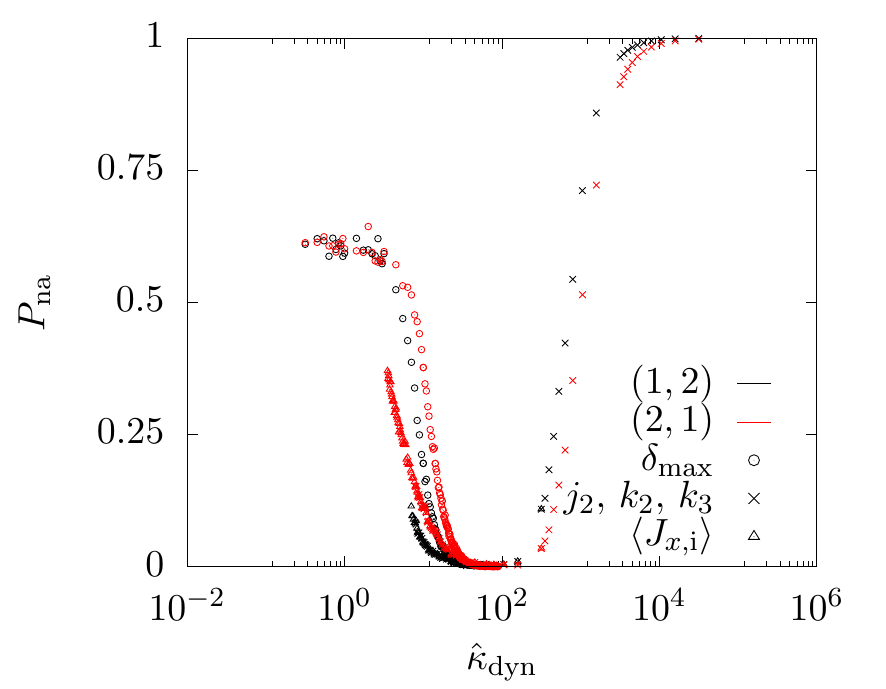}\\    
    \includegraphics[trim=5truemm 0truemm 5truemm 0truemm, width=.45\textwidth]{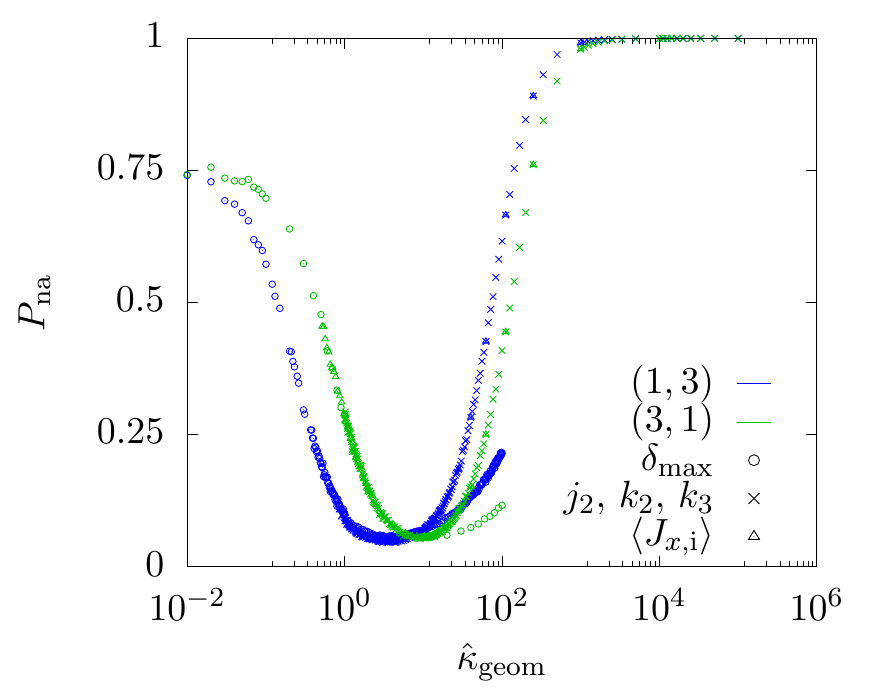}
    \hspace{10truemm}
    \includegraphics[trim=5truemm 0truemm 5truemm 0truemm, width=.45\textwidth]{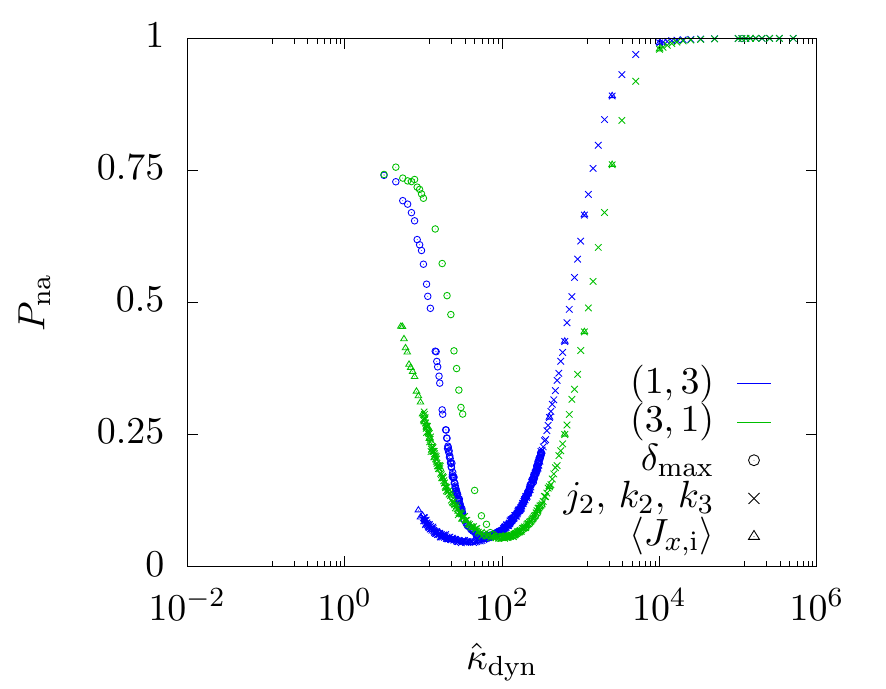}\\    
    \caption{Plots obtained by combining the numerical data presented in Fig.~\ref{fig:plot_deltaG} (left and right) and in Fig.~\ref{fig:plot_sigma} (left), using, as independent variable, $\hat\kappa_\text{geom}$, introduced in Eq.~\eqref{eq:kg_map}, for the left plots, and  $\hat\kappa_\text{dyn}$, from Eq.~\eqref{eq:kd_map}, for the right ones. The top plots refer to third-order while the bottom plots to fourth-order resonances. The colours encode the resonance considered, while the different point styles the variable that is varied in the data set, namely $\delta_\text{max}$, the strength of the non-linear magnets or the initial distribution width.}
    \label{fig:combine_plot}
\end{figure}

Figure~\ref{fig:plot_sigma} shows the dependence of $P_\text{na}$ on the initial emittance values and the ratio between vertical and horizontal emittances. On the left plot, we keep $\av{\Ji{x}}=\av{\Ji{y}}$ and we change their value, while on the right plot we keep $\av{\Ji{x}}=\num{1e-4}$ and we vary $\av{\Ji{y}}$ from $\num{1e-6}$ to $\num{1e-2}$. As observed before, the behaviour for the third- and fourth-order resonances are different. The first type of resonances features a virtually zero $P_\text{na}$ over a rather wide range of parameters under consideration. On the other hand, the fourth-order resonances feature a non-zero minimum for $P_\text{na}$, and that is achieved for well-defined values of the parameters under consideration. 

Furthermore, for all the four resonances $P_\text{na}$ increases (therefore that the emittance sharing is less effective) for large values of the initial emittances. This is due to a lower number of particle effectively performing the adiabatic jump. It also increases for small values of the initial action, as this translates to a more difficult onset of adiabatic conditions. It is also observed that achieving emittance sharing for fourth-order resonances is far more difficult than for the third-order ones when $\av{\Ji{y}} \neq \av{\Ji{x}}$. In particular, we remark that for $(1,\, 3)$ and $(3,\, 1)$ the situation is reversed, as emittance sharing fails for $(1,\, 3)$  when $\av{\Ji{y}}\ll\av{\Ji{x}}$, whereas this occurs for  $\av{\Ji{y}}\gg\av{\Ji{x}}$ in the case of the $(3,\, 1)$ resonance. This fact will be discussed later.

Note that, in Fig.~\ref{fig:plot_sigma} (right), some discontinuities are present. They are due to the initial condition, for which $\Ji{x}\approx n/m\,\Ji{y}$ and then  $P_\text{na}$ (see Eq.~\eqref{eq:pna}) has a small denominator and tends to diverge.

Some common observations can be drawn from Figs.~\ref{fig:plot_deltaG} and~\ref{fig:plot_sigma}. First of all, it is clear that, in general, third-order resonances achieve smaller values of $P_\text{na}$, than fourth-order ones. In the observed conditions, the best results for resonances $(1,\, 2)$ and $(2,\, 1)$ correspond to$P_\text{na}\approx0.01$, while for $(1,\, 3)$ and $(3,\, 1)$ the best performance corresponds to $P_\text{na}\approx 0.06$. This is explained by the fact, as can be seen from the higher values of $P_\text{na}$ at low initial action in the left plot of Fig.~\ref{fig:plot_sigma}, that fourth-order resonances are more affected than third-order ones by adiabaticity (note that the numerical simulations for fourth-order resonances were performed with a number of turns an order of magnitude higher than those of the third-order resonances).

In Fig.~\ref{fig:combine_plot}, we combine data from both plots of Fig.~\ref{fig:plot_deltaG} and the left plot of Fig.~\ref{fig:plot_sigma}, using as independent variables one of
\begin{subequations}
\begin{align}
\label{eq:kg_map}
\hat\kappa_\text{geom} & = \frac{\delta_\text{max}}{g \sqrt{\av{\Ji{x}}^{m+n-2}}} \\
\label{eq:kd_map}
\hat\kappa_\text{dyn} & = \frac{1}{g}\sqrt\frac{{\delta_\text{max}}}{{\av{\Ji{x}}^{m+n-2}}} \, ,
\end{align}
\end{subequations} 
%
%\begin{equation} 
%\hat\kappa_\text{dyn} = \frac{1}{g}\sqrt\frac{{\delta_\text{max}}}{{\av{\Ji{x}}^{m+n-2}}} \qquad \text{ in the right plot,} 
%\label{eq:kd_map}
%\end{equation}
%
where $g$ stands for the generic strength of the non-linear magnets, which, according to the resonance, is $k_2$, $j_2$ or $j_3$. The two new parameters differ from those introduced in Eqs.~\eqref{eq:kgeom} and \eqref{eq:kdyn} only for their adaptation to the configuration of the numerical simulations that feature  initial distributions where $\av{\Ji{x}}=\av{\Ji{y}}$. The goal of this analysis is to identify in which regime these global parameters are the relevant quantities to describe the emittance-sharing process: in that case, the data obtained by varying each parameter entering in the expression of the global parameters should lie on the same curve.

It is clearly visible that, when $\hat\kappa_\text{geom}$ and $\hat\kappa_\text{dyn}$ are small (\ie $\delta_\text{max}$ is small, the strength of the non-linear magnets is large, and the  distribution of initial conditions is wide), $P_\text{na}$ depends primarily on $\hat\kappa_\text{geom}$: the performance of the emittance sharing is only limited by the fact that the tune is varied only over a finite range.

The disagreement between the data collected while varying $\delta_\text{max}$ and the other quantities is visible when considering the resonance $(1,\, 2)$, but it can be understood by considering that when transforming the map~\eqref{eq:henon} with a normal sextupole to a resonant normal form, and truncating at second order in the action variables, contributions to the amplitude detuning are present, even in the absence of an octupolar term. Indeed, the coefficients $\alpha_{11}$ and $\alpha_{12}$ are proportional to $k_2^2$. As $\alpha_{12}\neq 0$, the separatrix crossing in $[\delta_\text{max},-\delta_\text{max}]$ is no longer symmetric, and a lower number of particles undergoes emittance sharing. 

For large values of the two parameters, $\hat\kappa_\text{dyn}$ captures the correct scaling, as in this regime, corresponding to large $\delta_\text{max}$, low $g$, and small initial emittance with constant value of $1/N$, the emittance sharing effectiveness is limited by the degree of adiabaticity of the process.

We remark that the scaling $g\av{\Ji{x}}^{(r-2)/2}$ always holds, since the strength of the non-linear magnets, for the model considered in this paper that features a single multipole, can always be normalised to unity under a convenient co-ordinate rescaling, therefore changing the average value of $\Ji{x}$.

\begin{figure}[htb]
    \centering
    \includegraphics[trim=5truemm 0truemm 5truemm 0truemm, width=.45\textwidth]{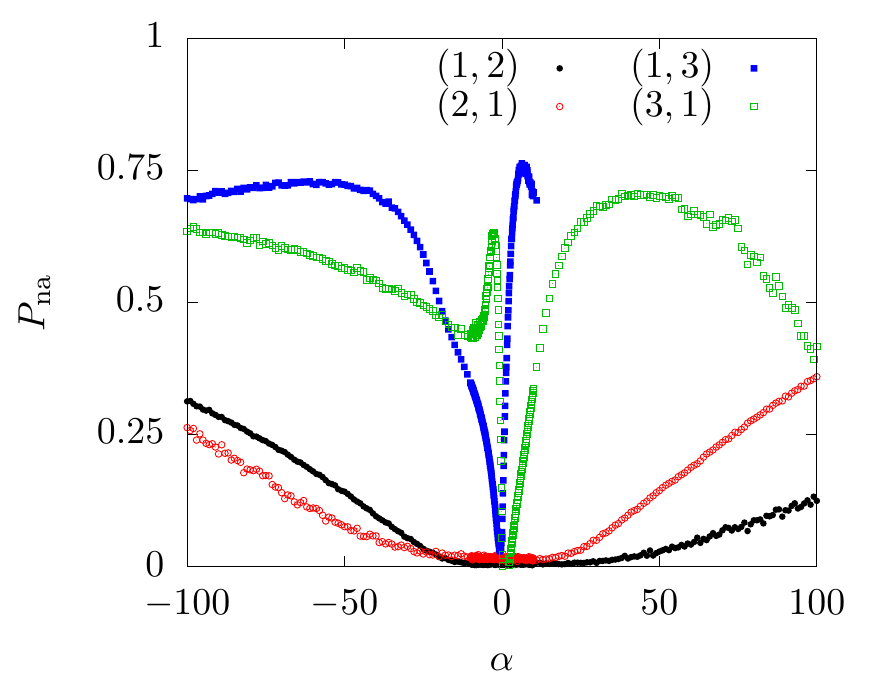}
    \hspace{10truemm}
    \includegraphics[trim=5truemm 0truemm 5truemm 0truemm, width=.45\textwidth]{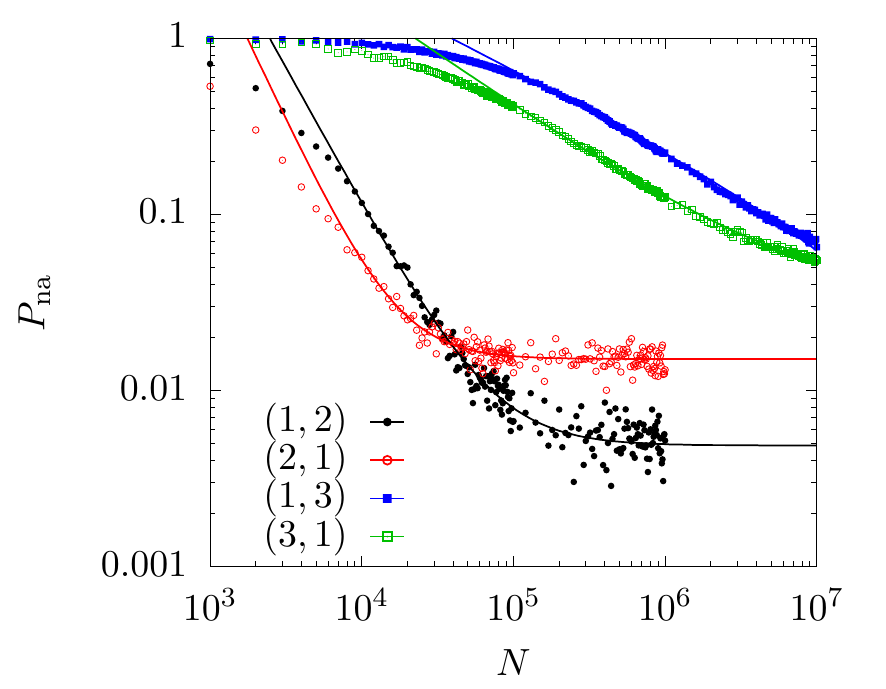}\\    
    \includegraphics[trim=5truemm 0truemm 5truemm 0truemm, width=.45\textwidth]{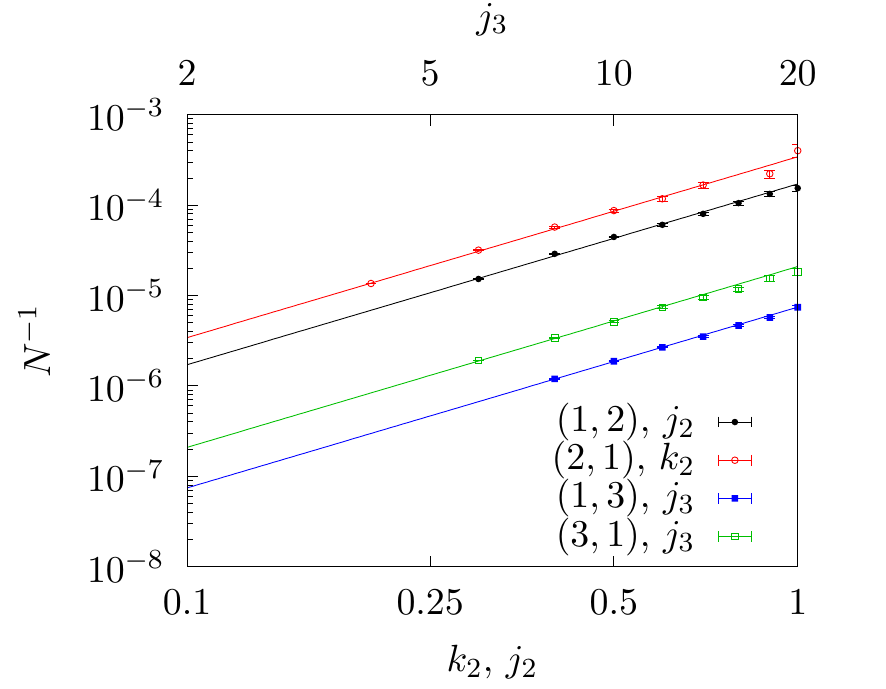}\\    
    \caption{Top-left: $P_\text{na}$ as a function of the amplitude-detuning parameter $\alpha$, for the four resonances. Top-right: $P_\text{na}$ as a function of the number of turns $N$. Power-law fits  $P_\text{na}=a_{m,n}N^{-b_{m,n}}+c_{m,n}$ are provided for each $(m,\, n)$ resonance. Bottom: the inverse of the number of turns for which $P_\text{na}$ reaches the value $P_\text{na}=0.2$ (for third-order resonances) and $P_\text{na}=0.3$ (for fourth-order ones) as a function of the strength of the non-linear magnets. Quadratic fit between $N^{-1}$ and the strength of the non-linear magnets are presented, confirming the scaling of Eq.~\eqref{eq:gscaling}. The parameters used for the plots are: $\av{\Ji{x}}=0.01$, $\delta_\text{max}=0.1$ for resonance $(1,2)$ and $(2,1)$ and $\delta_\text{max}=0.01$ for $(1,3)$ and $(3,1)$. For the top-left plot $N=10^6$, $k_2=1$ or $j_2=1$ (for third-order resonances), $N=10^7$, $j_3=10$ (for fourth-order ones) are used. The same values for the strength of the non-linear magnets are used in the to-right plot. Both the top-right and the bottoms plots use $\alpha=0$.}
    \label{fig:plot_alphan}
\end{figure}

In Fig.~\ref{fig:plot_alphan} (top-left) the role of  the amplitude-detuning parameter $\alpha$ is probed. Very different behaviours are observed depending on the resonance order. For the case of third-order resonances, a rather broad minimum of $P_\text{na}$ is observed around $\alpha=0$, which indicates that the presence of amplitude-detuning effects does not spoil the emittance sharing process. The situation is radically different for the case of the fourth-order resonances, where the presence of a non-zero amplitude detuning changes the number and the stability type of the fixed points of the systems under consideration. This is indicated by the presence of a very sharp minimum of $P_\text{na}$ around $\alpha=0$ with a steep increase in the close neighbourhood. 

In the top-right plot of Fig.~\ref{fig:plot_alphan}, the dependence of $P_\text{na}$ on the number of turns $N$ is shown. A fit using a power law $P_\text{na}=a_{m,n}N^{-b_{m,n}}+c_{m,n}$ provides an excellent agreement with the numerical data. This observation is crucial, as it reveals the intrinsic difference between the behaviour of the crossing of these non-linear 2D resonances with respect to that of the linear $(1,\, 1)$ resonance studied in Ref.~\cite{PhysRevAccelBeams.24.094002}. For the case of the linear coupling resonance, an exponential law for $P_\text{na}$ was found in the absence of amplitude detuning. The difference can be explained since the Hamiltonian describing the crossing of the linear coupling resonance is analytic, as the unstable fixed points in the action-angle coordinates are only a pathology of the coordinate system, while the Hamiltonian describing the 2D non-linear resonances features real unstable fixed points, The separatrices related with these unstable fixed points introduce an error in the action, which is the adiabatic invariant of the system, after the jump from one region to the other one of the phase space. Such a jump in the value of the adiabatic invariant has a power-law dependence on the number of turns. The values of the exponent of the fit law for $P_\text{na}$ are: $b_{1,2}= \num{1.56\pm .02}$, $b_{2,1}=\num{1.84\pm .08}$, $b_{1,3}=\num{0.43\pm .01}$ and $b_{3,1}=\num{0.60\pm 0.01}$, which reveal that they are strongly model dependent.

The bottom plot of Fig.~\ref{fig:plot_alphan} shows the scaling between the strength of the non-linear magnets and the inverse of the number $N$ of turns, which is a direct measurement of the adiabaticity of the emittance sharing process. The data show, for different values of $k_2$, $j_2$ and $j_3$ (depending on the resonance considered), the value of $N$ for which one obtains a small value for $P_\text{na}$, \ie $P_\text{na}=0.2$ for third-order resonances and $P_\text{na}=0.3$ for fourth-order ones. The curves represent quadratic functions that fit in an excellent way the numerical data, thus confirming the quadratic scaling found in Eq.~\eqref{eq:gscaling}, which is independent on $m$ and $n$.

\begin{figure}[p]
    \centering
    \includegraphics[height=.9\textheight]{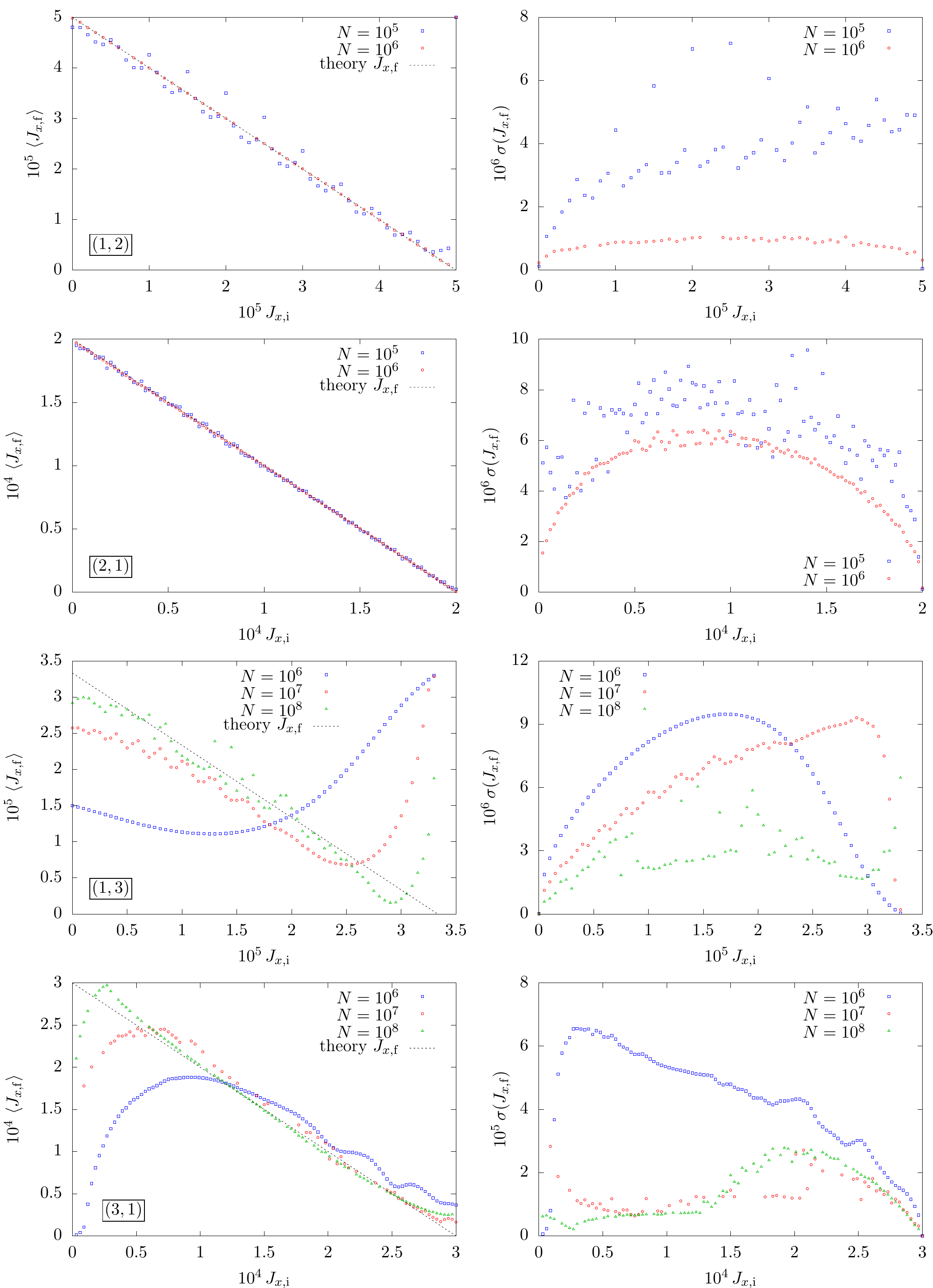}
    \caption{Average (left column) and variance (right column) of $\Jf{x}$, from initial uniform distributions at fixed $\Ji{x}$ chosen in the allowed interval $0\le \Ji{x} \le \Ji{2}/n$, having fixed $J_2=\num{1e-4}$. The average of $\Jf{x}$ is compared to the theoretical value $\Jf{x}=mJ_2/n -\Ji{x}$. The map~\eqref{eq:henon} has been used with no amplitude-detuning terms, using different numbers of turns $N$. For the third-order resonances, $\delta_\text{max}=0.1$, $j_2=1$ or $k_2=1$ have been used, whereas for the fourth-order ones  $\delta_\text{max}=0.01$, $j_3=10$ have been used.}
    \label{fig:plot_J}
\end{figure}

Finally, in Fig.~\ref{fig:plot_J} we analyse the emittance sharing process by generating uniform distributions of $10^4$ initial conditions at a fixed value of $\Ji{x}$, in the range $0\le \Ji{x} \le mJ_2/n$, having fixed $J_2=\num{1e-4}$. In the plots in the left column, we compare, for each resonance $(m,\, n)$, $\av{J_x}$ at the end of the process with the expected value from the theory, namely
\begin{equation}
    \Jf{x} = \frac{m}{n}J_2 - \Ji{x} \, ,
\end{equation}
computed for the case of a perfect emittance-sharing process. The results of numerical simulations are presented for different number of turns $N$ (for $N=\num{1e8}$, only $10^3$ initial conditions have been used due to constraints on the available CPU time). In the plots in the right column, the standard deviation of the values of $\Jf{x}$ is shown. The rows correspond to the various resonances considered. When increasing the number of turns, the average jump becomes closer to the theoretical expectation. For the resonances $(1,\, 2)$ and $(2,\, 1)$ we remark that the data oscillate around the expected value, in a similar fashion to what was observed in Ref.~\cite{PhysRevAccelBeams.24.094002} for an analogous situation with linear coupling. For the fourth-order resonances, the effects of the more complicated phase-space topology are clearly visible. For instance, resonance $(1,\, 3)$ suffers from a slow convergence of the data towards the expected values for large $\Ji{x}$, while for resonance $(3,\, 1)$ the same occurs, but rather at low values of $\Ji{x}$. This is consistent with what found when analysing initial conditions with $\av{\Ji{x}}\neq \av{\Ji{y}}$: resonance $(1,\, 3)$ fails at high values of $\Ji{x}$, while $(3,\, 1)$ at small ones. In the latter case, also, a variance bump is observed for $\Ji{x}>\num{1.5e-4}$ also at high adiabaticity: this is an effect of the presence of extra regions in the phase space. Finally, the plots of fourth-order resonances show how slowly the emittance sharing converges to the expected value when the number of turns is increased, which  explains the lower performance for emittance sharing for fourth-order resonances.

\section{Conclusions} \label{sec:conc}

In this paper, a novel beam manipulation technique is presented, based on the crossing of a 2D non-linear resonance to induce a sharing of the transverse emittances. The foundations of this technique have been discussed using Hamiltonian models and the adiabatic theory applied to resonance crossing. The performance of this manipulation has been assessed by means of detailed numerical simulations using map models, which are more realistic than the Hamiltonian ones. The results of the numerical simulations indicate that it is indeed possible to control the proposed process so to achieve a sharing of the transverse emittances. The final distributions of initial conditions retain the Gaussian character of the initial ones, which is an excellent feature. Scans of the various system parameters have been performed, thus achieving a good understanding of the details of the proposed mechanism. 

Differences in the behaviour and performance of the emittance-sharing process have been found and when comparing third- and fourth-order resonances, although these observations can be fully understood and explained in terms of the phase-space topology linked with each of the resonances under study.

As far as applications are concerned, this study shows clearly the theoretical feasibility of an emittance sharing process where the target emittance is met at up to $\approx 99\%$, using third-order resonances, and up to $\approx 90\%$, using fourth-order ones. These results are extremely encouraging, also in consideration of the fact that the resonances under consideration can be excited by widespread magnetic elements, such as normal and skew sextupoles, or skew octupoles (that can also be substituted by pairs of normal and skew sextupoles).

In summary, the novel beam manipulation passed successfully through theoretical and numerical tests and it is now ready for experimental validation.

\section{Acknowledgements}
We would like to express our warm gratitude to A.~Neishtadt for several discussions.
%
%\end{acknowledgement}
%
%\section*{Data availability}
%
%Data sharing not applicable to this article as no data sets were generated or analysed during the current study.
%
\clearpage

\appendix

%\section{Study of the fixed points of the Hamiltonian~\eqref{eq:ham_mn_J1J2}} \label{app:fixed_points}

\section{Magnet type and resonances: analysis using Normal Forms} \label{app:magnets}

The goal of this appendix is to compute, using Normal Form theory, which resonance can be excited by a certain nonlinearity in an H\'enon-like map as the one of Eq.~\eqref{eq:henon}.

We start the analysis by considering which monomial appears in the complex representation of the generic polynomial map. When dealing with 4D complex coordinates $(z_1,z_1^*,z_2,z_2^*)$, we use the vector notation $(\ell_1,m_1,\ell_2,m_2)$ to identify a monomial $ z_1^{\ell_1}\,{z_1^*}^{m_1}\,z_2^{\ell_2}\,{z_2^*}^{m_2}$, and we  indicate a 4D complex function as $\mathbf{F}=(F_1, F_1^*, F_2, F_2^*)$.

Starting from an H\'enon-like map, we replace the real variables with complex ones, defined according to $z_1 = x - ip_x$, $z_2 = y - ip_y$, together with the corresponding complex conjugate relationships, and we obtain, expanding all binomials, the following complex map
\begin{equation}
    \begin{split}
z_1' = e^{i\omega_1}\Bigg[ z_1 &+ \sqrt{\beta_x}  \frac{k_r\beta_x^\frac{r}{2}}{2^r r!}\sum_{q\le r/2}\sum_{\ell=0}^{r-2q}\sum_{p=0}^{2q} (-1)^q \left ( \frac{\beta_y}{\beta_x} \right )^q \binom{r}{2q}\binom{r-2q}{\ell} \times \\
& \times \binom{2q}{p} z_1^{r-2q-\ell}{z_1^*}^\ell z_2^{2q-p} {z_2^*}^p + \\
    &-\sqrt{\beta_y} \frac{j_r \beta_x^\frac{r}{2}}{2^r r!}\sum_{q\le(r-1)/2}\sum_{\ell=0}^{r-2q-1}\sum_{p=0}^{2q+1} (-1)^q \left ( \frac{\beta_y}{\beta_x} \right )^q \binom{r}{2q+1} \times\\
     & \times \binom{r-2q-1}{\ell}\binom{2q+1}{p} z_1^{r-2q-\ell-1}{z_1^*}^\ell z_2^{2q-p+1} {z_2^*}^p \Bigg] 
     \end{split}
\end{equation}
and
\begin{equation}
    \begin{split}
z_2' = e^{i\omega_2}\Bigg[ z_2 &- \frac{\beta_y}{\sqrt{\beta_x}}  \frac{k_r \beta_x^\frac{r}{2}}{2^r r!}\sum_{q\le (r-1)/2}\sum_{\ell=0}^{r-2q-1}\sum_{p=0}^{2q+1} (-1)^q \left ( \frac{\beta_y}{\beta_x} \right )^{q} \binom{r}{2q+1} \times\\
     & \times \binom{r-2q-1}{\ell}\binom{2q+1}{p} z_1^{r-2q-1-\ell}{z_1^*}^\ell z_2^{2q+1-p} {z_2^*}^p \\
    &- \sqrt{\beta_y} \frac{j_r \beta_x^\frac{n}{2}}{2^r r!}\sum_{q\le r/2}\sum_{\ell=0}^{r-2q}\sum_{p=0}^{2q} (-1)^q \left ( \frac{\beta_y}{\beta_x} \right )^{q}\binom{r}{2q}\binom{r-2q}{\ell} \times \\
    & \times \binom{2q}{p} z_1^{r-2q-\ell}{z_1^*}^\ell z_2^{2q-p} {z_2^*}^p \Bigg] \, ,
    \end{split}
\end{equation}
and we express the map action as $\mathbf{z}' = \mathbf{F}(\mathbf{z})$. $F_1$ includes the following monomials
\begin{itemize}
    \item for a normal multipole $k_r$, the generic term $(r-2q-\ell,\,\ell,\,2q-p,\,p)$ with
    \[
    0\le q\le \frac{r}{2}, \qquad 0\le \ell\le r-2q, \qquad 0\le p\le 2q \, ;
    \]
    \item for a skew multipole $j_r$, the generic term $(r-(2q+1)-\ell,\,\ell,\,2q+1-p,\,p)$ with
    \[
    0\le q\le \frac{r-1}{2}, \qquad 0\le \ell\le r-(2q+1), \qquad 0\le p\le 2q+1 \, ;
    \]
\end{itemize}
while for $F_2$ we obtain
\begin{itemize}
    \item for a normal multipole $k_r$, the generic term $(r-(2q+1)-\ell,\,\ell,\,2q+1-p,\,p)$ with
    \[
    0\le q\le \frac{r-1}{2}, \qquad 0\le \ell\le r-(2q+1), \qquad 0\le p\le 2q+1 \, ;
    \]
    \item for a  skew multipole $j_r$, the generic term $(r-2q-\ell,\,\ell,\,2q-p,\,p)$ with
    \[
    0\le q\le \frac{r}{2}, \qquad 0\le \ell\le r-2q, \qquad 0\le p\le 2q \, ,
    \]
\end{itemize}
and we remark that the form of the monomials for the normal and skew components exchanges between $F_1$ and $F_2$.

The Normal Form map $\mathbf{U}$ contains all the terms that are needed to solve the functional equation $\mathbf{F} \circ \bm{\Phi} = \bm{\Phi} \circ \mathbf{U}$, and, at order $k$, we have the functional equation
\[\Delta [\bm{\Phi}]_k(\bm{\zeta}) + [\mathbf{U}]_k(\bm{\zeta}) = [\mathbf{Q}]_k(\bm{\zeta})\]
where $[\mathbf{Q}]_k=[\mathbf{F} \circ \bm{\Phi}]_k$, the symbol $[ \cdot ]_k$ indicates the truncation of the homogeneous polynomial at order $k$, while $\mathbf{U}$ contains all monomials that are in the kernel of the Normal-Form operator $\Delta$, defined as $ \Delta \mathbf{\Phi}(\mathbf{\zeta})= e^{i \mathbf{\omega}} \mathbf{\Phi}(\mathbf{\zeta})-\mathbf{\Phi}(e^{i \mathbf{\omega}} \mathbf{\zeta})$, which,  following~\cite[p.~128]{Bazzani:262179}, occurs for monomials satisfying
\begin{equation}\begin{split} F_1: \qquad \omega_x (\ell_1-m_1) + \omega_y(\ell_2 - m_2) = \omega_x\,,\\
F_2: \qquad \omega_x (\ell_1-m_1) + \omega_y(\ell_2 - m_2) = \omega_y\,.
\end{split}\end{equation}
For a $(m,\, n)$ resonance (with $m$, $n$ coprimes), we can rewrite the previous relations as

\begin{subequations}
\begin{align}
\label{resF1}
     F_1: & \qquad \ell_1 - m_1 + \frac{m}{n}(\ell_2 - m_2) = 1 \, ,\\
\label{resF2}
     F_2: & \qquad \frac{n}{m}(\ell_1 - m_1) + \ell_2 - m_2 = 1 \, .
\end{align}
\label{resonance}
\end{subequations}

A condition for the resonant terms to be present in the interpolating Hamiltonian is that the corresponding monomials in the kernel of $\Delta$ are present in either $F_1$ or $F_2$ at order $r$.

To satisfy the condition for $F_1$, we need 

\begin{equation}
\ell_2 - m_2 \equiv 0 \pmod n \, ,  
\label{eq:mod_res}
\end{equation}
which for a normal multipole translates to
\begin{equation}
2(q-p) \equiv 0 \pmod n\, ,
\label{eq:mod_res_norm}
\end{equation}
while for a skew one to
\begin{equation}
2(q-p)+1 \equiv 0 \pmod n\,.
\label{eq:mod_res_skew}
\end{equation}

The condition is always satisfied when $\ell_2=m_2$ and in this case $\ell_1=m_1+1$ and the form of the corresponding monomial is
\begin{equation}
 \left (\frac{m+n}{2}-p,\, \frac{m+n}{2}-p-1,\, p,\, p \right ) \, ,\end{equation}
which is a valid solution only if $m$ and $n$ are both odd.

Solutions are also found whenever $\ell_2-m_2=s\,n$ with $s \neq 0$. For a normal multipole, $2(m-p)\le 2(m+p) \le r$ and we can restrict $s$ to $0< s \le (m-1)/n + 1$.

If $m < n$, then only $s=1$ is possible and if $n$ is odd, no other solution can be found. If $n$ is even, we have  $2(q-p) = n$ for which the resonant condition~\eqref{resF1} becomes $m-p-\ell=1$ and the form of the generic monomial reads $(-p,\,m-p-1,\,p+n,\,p)$, from which we can only choose $p=0$, so that the solution for $n$ even and $m < n$ reads 
\begin{equation}
(0,\,m-1,\,n,\,0) \, .    
\label{eq:cond1}
\end{equation}

If $m>n$, again two cases should be considered. If $n$ is odd, then $s$ must be even and the resonant condition~\eqref{resF1} is 
\begin{equation}
m+n+s(m-n)=2(1 +p+\ell)
\label{eq:pqs}
\end{equation}
and $m$ must be odd. Let $m=2m'+1$, $n=2n'+1$ and $s=2s'$. Substituting, we get the relation $p+\ell = m'+n'+2s'(m'-n')$, and $\ell_1$ becomes $\ell_1=m'(1-2s') + n'(1-2s')  - p + 1 - 2 s'$. Since $s'>0$, $\ell_1$ is always negative and no solution can be found.

If, on the other hand, $n$ is even, then $m$ has to be odd, and from Eq.~\eqref{eq:pqs}, $s$ is necessarily odd. Therefore, we substitute $n=2n'$, $m=2m'+1$, $s=2s'+1$ and we obtain $p+\ell =2m'+2m's'+s'-2n's'$
and $\ell_1 = -p-2n's'-2m's'-s'$. The only possibility, since $\ell_1\ge 0$, is given by $p=0$ and $s'=0$ (\ie $s=1$), which gives $\ell = 2m' = m-1$, $2q= n$, and the general form of the resonant monomial is therefore 
\begin{equation}
(0,\,m-1,\,n,\,0) \, ,    
\end{equation}
which is the same as for the previous case~\eqref{eq:cond1}

We proceed in a similar fashion in the case of a skew multipole. The first point to note is that the special solution of Eq.~\eqref{eq:mod_res} for $s=0$ does not exist in this case. Hence, the resonant condition~\eqref{eq:mod_res_skew} $2q-2p+1=s\,n$ provides a solution only if both $s$ and $n$ are odd. Hence, letting $n=2n'+1$ and $s=2s'+1$ and performing the usual substitutions, we get $q = 2s'n'+s'+n'+p$ and $\ell = m-2s'n'-s'-p-1+s'm$ and the condition $\ell_1\ge 0$ becomes $-2s'n'-p-s'm\ge 0$, which requires $p=0$ and $s'=0$, and the generic skew monomial is of the form
\begin{equation}
(0,\,m-1,\,n,\,0) \, .
\end{equation}

In summary, the monomials of $U_1$ are:
\begin{equation} 
\begin{aligned}
\text{Normal multipoles:} &
\quad\begin{cases} \displaystyle{\qty(\frac{m+n}{2}-p,\,\frac{m+n}{2}-p-1,\,p,\,p)},\quad 0\le p \le \frac{r-1}{2} &\qquad \text{if} \quad m, n \; \text{are odd,} \\
\\
(0,\,m-1,\,n,\, 0) &\qquad \text{if} \quad n \; \text{is even} \,,
\end{cases} \\
\text{Skew multipoles:} & \quad
%\begin{cases} 
(0,\,m-1,\,n,\, 0) \qquad \text{if} \quad n \; \text{is odd.}
%\end{cases}
\end{aligned}
\end{equation}
%For the sake of completeness, the degenerate case $p=q$ is considered here. In this case $n=2p-1$ and, most importantly, the solution of Eq.~\eqref{resF1} should not satisfy any divisibility condition the only condition imposed is

%\[ q-\ell-m=1 \]

%no matter the type of multipole considered. Therefore, the following resonant monomials arise

%\begin{align}
%\text{normal}: \quad (2q-2q-\ell-1, \,\ell, \, 2q+\ell+1-q,\, q-\ell-1) & \\ 
% \text{for} \quad \ell \leq q-1, \quad q-1-\ell \leq & 2q \leq 2q-\ell -1 \nonumber \\
%\text{skew}: \quad (2q-2q-\ell-2, \,\ell, \, 2q+\ell+2-q,\, q-\ell-1) & %\\  
% \text{for} \quad \ell \leq q-1, \quad q-2-\ell \leq & 2q \leq 2q-\ell -2 \nonumber
% \label{degenerate}
%\end{align}

Thus far, only Eq.~\eqref{resF1} has been considered. For $F_2$, we proceed in the same way. We need to apply the resonant condition~\eqref{resF2}, which requires $\ell_1-m_1 \equiv 0 \pmod m$, \ie
\begin{equation}
\frac{n}{m}(\ell_1-m_1)=s\, n\,.
\end{equation}

For a normal multipole, the divisibility condition and the resonant correspond to
\begin{equation} 
\begin{cases} m(1-s) + n &= 2(q + \ell + 1)\\
2(q-p) &= s\,n\,,
\end{cases}
\end{equation}
and solving for $\ell$ and $q$, and substituting in $\ell_1$ we obtain $\ell_1 = \frac{1}{2}(m+n)(1+s) - 1 - p$. We have $0\le\ell_1\le r=m+n-1$, because $(m+n)(1+s) \ge 2(p+1) \ge 0$ implies $s\le 0$, and $(m+n)(1+s) \le 2(r+1+p)$ implies $1+s \le 2\qty( 1 + r/(r-1)) \le 2$, therefore $s\le 1$. This means that we can restrict $s$ to the two values $s=0$ or $s=1$.

For $s=0$, we obtain the solution
\begin{equation}
\qty(\frac{m+n}{2}-p-1,\,\frac{m+n}{2}-p-1,\,p+1,\,p)\qquad \text{for} \quad 0\le p\le r/2\,,
\end{equation}
which only exists if $m+n$ is even, so both $m$ and $n$ must be odd, while, for $s=1$
\begin{equation}
(m+n-p-1,\,n-p-1,\,p-n+1,\, p)\,.
\end{equation}

The only way to ensure both $m_1\ge 0$ and $\ell_2\ge 0$ is setting $m_1=\ell_2=0$, so $p=n-1$, whence we obtain the resonant solution 
\begin{equation}
(m,\,0,\,0,\, n-1)
\end{equation}

In the case of a skew multipole, we have the equations 
\begin{equation} 
\begin{cases} m+n-1-2q-2\ell &= sm\\
2q &= 1 + 2p - sn
\end{cases}
\end{equation}
and the same condition for $\ell_1$, which imposes as before, $s=0$ or $s=1$. For $s=0$, we have the same solution (if $m$ and $n$ are odd)
\begin{equation}\qty(\frac{m+n}{2}-p-1,\,\frac{m+n}{2}-p-1,\,p+1,\,p)\,,\end{equation}
while, for $s=1$, as before
\begin{equation}
(m,\,0,\,0,\, n-1)\,.
\end{equation}

Now, if we substitute $\ell_1=m$, $m_1=\ell_2=0$ and $m_2=n-1$ in the generic Normal Form term in $U_2$, we need $2q=n-2$, so $q = (n-2)/2$, which is possible only if $n$ is even. In the skew case, on the other hand, we have $2q=n-1$, which is solved only for odd $n$. Thus, summarising:
\begin{equation} 
\text{Normal multipoles:} 
\begin{cases} \displaystyle{\qty(\frac{m+n}{2}-p-1,\,\frac{m+n}{2}-p-1,\,p+1,\,p)} \qquad \text{if} \quad m, n \; \text{are odd} \\
\\
(m,\,0,\,0,\, n-1) \qquad \text{if} \quad n \; \text{is even} \,,
\end{cases}
\end{equation}
\begin{equation} 
\text{Skew multipoles:} 
\begin{cases} \displaystyle{\qty(\frac{m+n}{2}-p-1,\,\frac{m+n}{2}-p-1,\,p+1,\,p)} \qquad \text{if} \quad m, n \; \text{are odd} \\
\\
(m,\,0,\,0,\, n-1) \qquad \text{if} \quad n \; \text{is odd} \,,
\end{cases}
\end{equation}

%\begin{itemize}
%    \item Normal multipoles: $\qty(\frac{m+n}{2}-p-1,\,\frac{m+n}{2}-p-1,\,p+1,\,p)$ if $m$ and $n$ are odd, $(m,\,0,\,0,\, n-1)$ if $n$ is even;
%    \item Skew multipoles: $\qty(\frac{m+n}{2}-p-1,\,\frac{m+n}{2}-p-1,\,p+1,\,p)$ if $m$ and $n$ are odd, $(m,\,0,\,0,\, n-1)$ if $n$ is odd.
%\end{itemize}

The next step is to consider how the resonant monomials in the Normal Form $\mathbf{U}$ contribute to the interpolating Hamiltonian. The first point consists in showing that monomials of type
\begin{equation} \left (\frac{m+n}{2} - p,\, \frac{m+n}{2} - p-1,\, p,\,p \right ) \end{equation}
do not contribute to the resonant part of the interpolating Hamiltonian. In fact, given a 4D Normal Form $\vb{U}(\bm\zeta,\bm\zeta^*)=(U_1,U^*_1,U_2,U^*_2)$, and writing  polynomials as:
\begin{equation}
A(\zeta_1,\zeta_1^*,\zeta_2,\zeta_2^*) = \sum_{\ell_1,\,m_1,\,\ell_2,\,m_2} A(\ell_1,\,m_1,\,\ell_2,\,m_2)\zeta_1^{\ell_1}{\zeta_1^*}^{m_1}\zeta_2^{\ell_2}{\zeta_2^*}^{m_2}\,,
\end{equation}
the construction of the interpolating Hamiltonian of order $r$ is done as follows:
\begin{align*}
\mathcal{H}(\ell_1+1,\,m_1,\,\ell_2,\,m_2) &= -\frac{1}{\ell_1+1}U_1^*(\ell_1,\,m_1,\,\ell_2,\,m_2) &\qquad \text{for } \ell_1+m_1+\ell_2+m_2=r\,,\\
\mathcal{H}(0,\,m_1+1,\,\ell_2,\,m_2) &= \frac{1}{m_1+1}U_1(0,\,m_1,\,\ell_2,\,m_2) &\qquad \text{for } m_1+\ell_2+m_2=r\,,\\
\mathcal{H}(0,\,0,\,\ell_2+1,\,m_2) &= -\frac{1}{\ell_2+1}U_2^*(0,\,0,\,\ell_2,\,m_2) &\qquad \text{for } \ell_2+m_2=r\,,\\
\mathcal{H}(0,\,0,\,0,\,r+1) &= \frac{1}{r+1}U_2(0,\,0,\,0,\,r)\,.
\end{align*}

The monomials in $U_1^*$ and $U_2^*$ are the same in $U_1$ and $U_2$, but with the exchanges $\ell_1\leftrightarrow m_1$ and $\ell_2 \leftrightarrow m_2$. Hence, the $U_1$ terms can be $(0,\,m-1,\,n,\,0)$ or $(\ell_2+1,\,\ell_2,\,m_1,\,m_1)$ and in $U^*_1$ we will have either $(m-1,\,0,\,0,\,n)$ or $(\ell_2,\,\ell_2+1,\,m_1,\,m_1)$. If this latter form is present in $U^*_1$, it gives rise to the Hamiltonian coefficient $\mathcal{H}(\ell_2+1,\,\ell_2+1,\,m_1,\,m_1)$, and by performing the transformation to the action-angle coordinates $(\bm J,\,\bm\phi)$, \ie $\zeta_1 = \sqrt{J_x}e^{i\phi_x}$, $\zeta_2 = \sqrt{J_y}e^{i\phi_y}$, the angular parts of these terms vanish, and we obtain a $J_x^{\ell_2/2}J_y^{m_1/2}$ monomial, which is clearly non resonant. 

For what concerns the second component of the map, we see from the construction of the interpolating Hamiltonian that we need to restrict our search to monomials with $\ell_1=m_1=0$. We can have terms in $U_2$ of the form 
\begin{equation}
\qty(0,\,0,\,\frac{r+1}{2},\,\frac{r-1}{2})
\end{equation}
owing to the resonant condition $\ell_2-m_2=1$ with $\ell_2+m_2=r$. Since such a term contributes to the Hamiltonian via $U_2^*$, it gives rise to the Hamiltonian term $\mathcal{H}(0,0,(r-1)/2 + 1, (r+1)/2) = \mathcal{H}(0,0,(r+1)/2, (r+1)/2)$, and, as we discussed before, this does not give origin to a resonant term since the angular parts of $\zeta_2$ and $\zeta_2^*$ are cancelled. Finally, we could look for monomials in $U_2$ of the form $(0,\,0,\,0,\,r)$, but the resonant condition would be $-r=1$ which is  never satisfied.

In conclusion, we have the following relationships between the parity of $m$ and $n$ and the type of multipole element that gives origin to quasi-resonant Hamiltonian terms:
\begin{center}
\begin{tabular}{ c|c|c|c|c } 
 Multipole type & $m$ & $n$ & $U_1$ non-trivial monomials & $\mathcal{H}$ resonant monomials\\
 \hline
\multirow{4}{*}{Normal}  & even & odd  & no & no \\
  & odd & odd  & no & no \\
  & odd & even  & $(0,\,m-1,\,n,\,0)$ & $(0,\,m,\,n,\,0);\,(m,\,0,\,0,\,n)$\\
 \hline
\multirow{4}{*}{Skew}  & even & odd  & $(0,\, m-1,\, n,\,0)$ & $(0,\,m,\,n,\,0);\,(m,\,0,\,0,\,n)$\\ 
  & odd & odd  &  $(0,\, m-1,\, n,\,0)$ & $(0,\,m,\,n,\,0);\,(m,\,0,\,0,\,n)$\\
  & odd & even  & no & no \\
 \end{tabular}
\end{center}

Therefore, if we restrict our search to sextupoles and octupoles ($r=2$ or $r=3$) we therefore find that a normal sextupole can excite the $(1,2)$ resonance, while a skew sextupole the $(2,1)$ one. A skew octupole, on the other hand, is needed for both the $(1,3)$ and $(3,1)$ resonances.

%In these conditions, we will be able to write a resonant Hamiltonian in the $(J_x,\,\phi_x,\,J_y,\,\phi_y)$ coordinates, and, coupling the $x$ and $y$ coordinates and averaging, reduce the Hamiltonian to the usual resonant form in $J_1$ and $\phi_1$.

\section{Computation of \texorpdfstring{$P_\text{na}$}{Pna} in resonant conditions} \label{app:rescond}

The motion in the resonant condition is governed by the Hamiltonian~\eqref{eq:ham_mn_J1J2} with $\delta=0$ (for the sake of simplicity, we neglect the amplitude-detuning terms in the following considerations). The analysis of the phase space topology shows that, independently of the resonance order, the allowed circle is symmetrically divided in two regions by the coupling arc. The trajectory of a particle with initial condition $(\phii{1},\Ji{1})$, in one of the two hemicircles (let us choose $\cos\phi>0$) is given by the solution of the equation
$\ham(\phi_1,J_1)=\ham(\phii{1},\Ji{1})$, \ie
\begin{equation}
    J_1^{m/2}(J_2-nJ_1)^{n/2}\cos\phi_1 =  \Ji{1}^{m/2}(J_2-n\Ji{1})^{n/2}\cos\phii{1} 
    \label{eq:rcond}
\end{equation}
whose solution gives rather straightforwardly the function $\phi_1(J_1)$ given the initial conditions.

To compute $P_\text{na}$ for an emittance sharing process when $\delta_\text{max}\to 0$, ultimately, we should consider a motion when $\delta$ is equal to zero, and since $P_\text{na}$ depends on $\av{\Ji{x}}$, our goal is to compute the trajectory $J_1(\phi_1)$. 

For generic values of $(m,\, n)$ one cannot easily invert analytically $\phi_1(J_1)$ from Eq.~\eqref{eq:rcond}, yet this task can be carried out numerically. To compute the final mean $J_1$ for a given initial distribution, we can use the time average of $J_1$ over a (long) time interval $T$. This is given by
\begin{equation}
    \overline{J_1} = \frac{1}{T}\int_0^T \dd t\, J_1(\phi_1) = \frac{1}{T}\int_{\phi_-}^{\phi_+}\dd\phi_1\, \frac{J_1(\phi_1)}{\dot{\phi}_1} = \frac{\displaystyle{\int_{\phi_-}^{\phi_+}\dd\phi_1\, \qty(J_1\qty(\phi_1)/\dot{\phi}_1)}}{\displaystyle{\int_{\phi_-}^{\phi_+}\dd\phi_1\, (1/\dot{\phi}_1)}},\,
\end{equation}
where $\phi_\pm$ are the inversion points of the trajectory and $\dot\phi_1$ is taken from the equations of motion with $\delta=0$. Note that the strength $G$ of the resonant term never appears in the integral.

Then, the averaging of the result of $\overline{J_1}$ over the initial conditions $(\phii{1}, \Ji{1})$ provides the expected value of $\av{\Ji{x}}$ as $\delta\to 0$, and therefore of $P_\text{na}$. 

For resonance $(1,2)$, if the Cartesian coordinate $X$ is used instead of the angle $\phi_1$ to parametrise the motion, $J_1(X)$ can be written as
\begin{equation}
    J_1 = \frac{1}{2}\qty(J_2 - \frac{X_\text{i}(J_2 - X_\text{i}^2)}{X})
\end{equation}
and the time average $\overline{J_1}$
\begin{equation} \overline{J_1} = \frac{J_2}{2} + \frac{C_\text{i}}{2}\frac{\displaystyle{\int_{X_-}^{X_+}\dd X\, \frac{1}{X^2\sqrt{J_2-X^2+C_\text{i}/X}}} }{\displaystyle{\int_{X_-}^{X_+}\dd X\, \frac{1}{X\sqrt{J_2-X^2+C_\text{i}/X}}}}
\end{equation}
where $C_\text{i}=X_\text{i}\sqrt{X_\text{i}^2-J_2}$ and the inversion points are $X_-=X_\text{i}=\sqrt{2\Ji{1}}\cos\phii{1}$, $X_+ = -(X_\text{i}+\sqrt{4J_2 - 3X_\text{i}^2})/2$. 

A numerical evaluation of this integral, averaged on a Gaussian distribution for $(\phii{1},\Ji{1})$ with $\av{\Ji{x}}=\num{1e-4}$ gives $\overline{J_1}=\num{8.115e-5}$, which replaced into the definition of $P_\text{na}$ gives a values, when $\delta\to 0$
\begin{equation}
    P_\text{na}=\num{.623},\,
\end{equation}
which is consistent with the value observed in Fig.~\ref{fig:plot_deltaG} (left). This procedure can be used to explain the values observed for other resonances, too.

% BibTeX users please use
\bibliographystyle{unsrt}
\bibliography{mybibliography}

\providecommand{\noopsort}[1]{}\providecommand{\singleletter}[1]{#1}%
\begin{thebibliography}{10}

\bibitem{Bazzani9948}
A.~Bazzani, S.~Siboni, and G.~Turchetti.
\newblock Diffusion in hamiltonian systems with a small stochastic
  perturbation.
\newblock {\em Physica D: Nonlinear Phenomena}, 76(1):8--21, 1994.

\bibitem{Kandrup1999DiffusionAS}
H.E. Kandrup, C.~Siopis, G.~Contopoulos, and R.~Dvorak.
\newblock Diffusion and scaling in escapes from two-degrees-of-freedom
  hamiltonian systems.
\newblock {\em Chaos}, 9 2:381--392, 1999.

\bibitem{NEISHTADT2006158}
A.I. Neishtadt and A.A. Vasiliev.
\newblock Destruction of adiabatic invariance at resonances in slow--fast
  hamiltonian systems.
\newblock {\em Nucl. Instrum. Meth. A}, 561(2):158--165, 2006.
\newblock Proceedings of the Workshop on High Intensity Beam Dynamics.

\bibitem{Neishtadt_2019}
A.I. Neishtadt.
\newblock On mechanisms of destruction of adiabatic invariance in
  slow{\textendash}fast hamiltonian systems.
\newblock {\em Nonlinearity}, 32(11):R53--R76, oct 2019.

\bibitem{Nekhoroshev:1971aa}
N.~Nekhoroshev.
\newblock Behavior of hamiltonian systems close to integrable.
\newblock In {\em Functional Analysis and Its Applications}, volume~5, page
  338. Kluwer Academic Publishers-Plenum Publishers, 1971.

\bibitem{Courant:593259}
E.D. Courant and H.S. Snyder.
\newblock {Theory of the Alternating-Gradient Synchrotron}.
\newblock {\em Ann. Phys.}, 3:1--48, 1958.

\bibitem{Bazzani:262179}
A.~Bazzani, G.~Servizi, E.~Todesco, and G.~Turchetti.
\newblock {\em {A normal form approach to the theory of nonlinear betatronic
  motion}}.
\newblock {CERN} Yellow Reports: Monographs. {CERN}, Geneva, 1994.

\bibitem{PhysRevLett.88.104801}
R.~Cappi and M.~Giovannozzi.
\newblock Novel method for multiturn extraction: Trapping charged particles in
  islands of phase space.
\newblock {\em Phys. Rev. Lett.}, 88:104801, 2002.

\bibitem{PhysRevSTAB.7.024001}
R.~Cappi and M.~Giovannozzi.
\newblock Multiturn extraction and injection by means of adiabatic capture in
  stable islands of phase space.
\newblock {\em Phys. Rev. ST Accel. Beams}, 7:024001, 2004.

\bibitem{Giovannozzi:987493}
M.~Giovannozzi, M.J. Barnes, O.E. Berrig, A.~Beuret, J.~Borburgh, P.~Bourquin,
  R.~Brown, J.-P. Burnet, F.~Caspers, J.-M. Cravero, T.~Dobers, T.~Fowler, S.S.
  Gilardoni, M.~Hourican, W.~Kalbreier, T.~Kroyer, F.~Di~Maio, M.~Martini,
  V.~Mertens, E.~Métral, K.D. Metzmacher, C.~Rossi, J.-P. Royer, L.~Sermeus,
  R.~Steerenberg, G.~Villiger, and T.~Zickler.
\newblock {\em {The {CERN} {PS} multi-turn extraction based on beam splittting
  in stable islands of transverse phase space: Design Report}}.
\newblock {CERN} Yellow Reports: Monographs. {CERN}, Geneva, 2006.

\bibitem{PhysRevSTAB.12.014001}
A.~Franchi, S.~Gilardoni, and M.~Giovannozzi.
\newblock Progresses in the studies of adiabatic splitting of charged particle
  beams by crossing nonlinear resonances.
\newblock {\em Phys. Rev. ST Accel. Beams}, 12:014001, 2009.

\bibitem{Borburgh:2137954}
J.~Borburgh, S.~Damjanovic, S.~Gilardoni, M.~Giovannozzi, C.~Hernalsteens,
  M.~Hourican, A.~Huschauer, K.~Kahle, G.~Le~Godec, O.~Michels, and
  G.~Sterbini.
\newblock {First implementation of transversely split proton beams in the
  {CERN} Proton Synchrotron for the fixed-target physics programme}.
\newblock {\em EPL}, 113(3):34001. 6 p, 2016.

\bibitem{PhysRevAccelBeams.20.014001}
S.~Abernethy, A.~Akroh, H.~Bartosik, A.~Blas, T.~Bohl, S.~Cettour-Cave,
  K.~Cornelis, H.~Damerau, S.~Gilardoni, M.~Giovannozzi, C.~Hernalsteens,
  A.~Huschauer, V.~Kain, D.~Manglunki, G.~M\'etral, B.~Mikulec, B.~Salvant,
  J.-L. Sanchez~Alvarez, R.~Steerenberg, G.~Sterbini, and Y.~Wu.
\newblock Operational performance of the {CERN} injector complex with
  transversely split beams.
\newblock {\em Phys. Rev. Accel. Beams}, 20:014001, 2017.

\bibitem{PhysRevAccelBeams.20.061001}
A.~Huschauer, A.~Blas, J.~Borburgh, S.~Damjanovic, S.~Gilardoni,
  M.~Giovannozzi, M.~Hourican, K.~Kahle, G.~Le~Godec, O.~Michels, G.~Sterbini,
  and C.~Hernalsteens.
\newblock Transverse beam splitting made operational: Key features of the
  multiturn extraction at the {CERN Proton Synchrotron}.
\newblock {\em Phys. Rev. Accel. Beams}, 20:061001, 2017.

\bibitem{PhysRevAccelBeams.22.104002}
A.~Huschauer, H.~Bartosik, S.~Cettour Cave, M.~Coly, D.~Cotte, H.~Damerau,
  G.~P. Di~Giovanni, S.~Gilardoni, M.~Giovannozzi, V.~Kain,
  E.~Koukovini-Platia, B.~Mikulec, G.~Sterbini, and F.~Tecker.
\newblock {Advancing the {CERN} proton synchrotron multiturn extraction towards
  the high-intensity proton beams frontier}.
\newblock {\em Phys. Rev. Accel. Beams}, 22:104002, Oct 2019.

\bibitem{Metral:529690}
E.~M\'etral.
\newblock {Simple theory of emittance sharing and exchange due to linear
  betatron coupling}.
\newblock Technical Report {CERN}-{PS}-2001-066-AE, {CERN}, Geneva, 2001.

\bibitem{PhysRevSTAB.10.064003}
A.~Franchi, E.~M\'etral, and R.~Tom\'as.
\newblock Emittance sharing and exchange driven by linear betatron coupling in
  circular accelerators.
\newblock {\em Phys. Rev. ST Accel. Beams}, 10:064003, Jun 2007.

\bibitem{PhysRevAccelBeams.23.044003}
M.~Aiba and J.~Kallestrup.
\newblock Theory of emittance exchange through coupling resonance crossing.
\newblock {\em Phys. Rev. Accel. Beams}, 23:044003, Apr 2020.

\bibitem{PhysRevAccelBeams.24.094002}
A.~Bazzani, F.~Capoani, M.~Giovannozzi, and A.~I. Neishtadt.
\newblock Adiabaticity of emittance exchange due to crossing of the coupling
  resonance.
\newblock {\em Phys. Rev. Accel. Beams}, 24:094002, Sep 2021.

\bibitem{PhysRevLett.110.094801}
S.~Y. Lee, K.~Y. Ng, H.~Liu, and H.~C. Chao.
\newblock Evolution of beam distribution in crossing a {Walkinshaw} resonance.
\newblock {\em Phys. Rev. Lett.}, 110:094801, 2013.

\bibitem{chao2015emittance}
Hung-Chun Chao.
\newblock {\em Emittance evolution in crossing Walkinshaw resonance and
  envelope dynamics simulations}.
\newblock PhD thesis, Indiana University, 2015.

\bibitem{kallestrup:ipac2021-mopab019}
J.~Kallestrup and X.~Gu.
\newblock {Possible Application of Round-to-Flat Hadron Beam Creation Using 3rd
  Order Coupling Resonances for the Electron-Ion Collider}.
\newblock In {\em Proc. IPAC'21}, number~12 in International Particle
  Accelerator Conference, pages 99--102. JACoW Publishing, Geneva, Switzerland,
  08 2021.
\newblock https://doi.org/10.18429/JACoW-IPAC2021-MOPAB019.

\bibitem{henon}
M.~H\'enon.
\newblock {Numerical study of quadratic area-preserving mappings}.
\newblock {\em Q. Appl. Math.}, 27:291, 1969.

\bibitem{Bazzani:1995vj}
A.~Bazzani, M.~Giovannozzi, and E.~Todesco.
\newblock {A Program to compute Birkhoff normal forms of symplectic maps in
  {$\mathbb{R}^4$}}.
\newblock {\em Comput. Phys. Commun.}, 86:199--207, 1995.

\bibitem{Arnold:937549}
V.I. Arnol'd, V.V. Kozlov, and A.I. Neishtadt.
\newblock {\em {Mathematical aspects of classical and celestial mechanics.
  Dynamical systems III; 3rd rev. version}}.
\newblock Encyclopaedia of mathematical sciences. Springer, Heidelberg, 2006.

\bibitem{neish1975}
A.I. Neishtadt.
\newblock Passage through a separatrix in a resonance problem with a
  slowly-varying parameter.
\newblock {\em Journal of Applied Mathematics and Mechanics}, 39(4):594 ---
  605, 1975.

\end{thebibliography}

%\begin{thebibliography}{}
%Author, \textit{Book title} (Publisher, place year) page numbers
% etc
%\end{thebibliography}

\end{document}